\documentclass[11pt]{article}

\usepackage[final]{acl}

\usepackage{times}
\usepackage{latexsym}

\usepackage[T1]{fontenc}

\usepackage[utf8]{inputenc}

\usepackage{microtype}

\usepackage{inconsolata}

\usepackage{graphicx}

\usepackage[utf8]{inputenc} 
\usepackage[T1]{fontenc}    
\usepackage{hyperref}       
\usepackage{url}            
\usepackage{booktabs}       
\usepackage{amsfonts}       
\usepackage{nicefrac}       
\usepackage{microtype}      
\usepackage{xcolor}         
\usepackage{graphicx}
\usepackage{amsmath}
\usepackage{float}
\usepackage{multirow}

\usepackage{booktabs}
\usepackage[table]{xcolor}
\usepackage{array}

\usepackage{color-edits}
\addauthor{vc}{blue}
\addauthor{jp}{magenta}
\addauthor{bh}{cyan}
\addauthor{sh}{red}
\addauthor{kx}{orange}

\title{Is Agent Code Less Maintainable Than Human Code?}

\author{
 \textbf{Shaswat Patel\textsuperscript{1}}\thanks{Equal contribution. \\ Code and data available at: \url{https://github.com/shaswatpatel123/CodeThread}},
 \textbf{Betty Li Hou\textsuperscript{1}}\footnotemark[1],
 \textbf{Arun Purohit\textsuperscript{1}},
 \textbf{Kai Xu\textsuperscript{1}},
 \textbf{Jane Pan\textsuperscript{1}},
 \textbf{He He\textsuperscript{1}},
 \textbf{Valerie Chen\textsuperscript{2}}
\\
 \textsuperscript{1}New York University
 \textsuperscript{2}Carnegie Mellon University\\
\texttt{\{spp9399, betty.li.hou\}@nyu.edu}
}

\begin{document}
\maketitle
\begin{abstract}
Maintainability is a core dimension of software engineering, shaping how code is written, reviewed, and developed over time. While coding agents have demonstrated strong performance on single-issue tasks, it remains unclear how maintainable their code is when future agents build on top of it, potentially leading to compounding downstream effects. We investigate how agent code compares to human code in these maintenance settings, presenting \textbf{CodeThread}, a framework to construct controlled experiments from repository-level coding benchmarks. Applying CodeThread to four frontier coding agents and four benchmarks, we find that agents are less effective at resolving tasks when building on agent code compared to human code, with task resolve rate drops of up to 13.1\%. Regression analysis reveals that many traditional software engineering maintainability metrics do not explain this difference. Instead, the clearest signals are subtler behavioral differences in agent code, such as changes to input validation and error handling, along with differences in downstream code size and task difficulty. These findings highlight the need to evaluate coding agents not only by immediate task resolution but also by maintainability, and point to potential sources of downstream errors introduced by agent code.
\end{abstract}

\section{Introduction}
\label{sec:intro}

\begin{figure*}[t]
    \centering
    \includegraphics[width=0.90\linewidth]{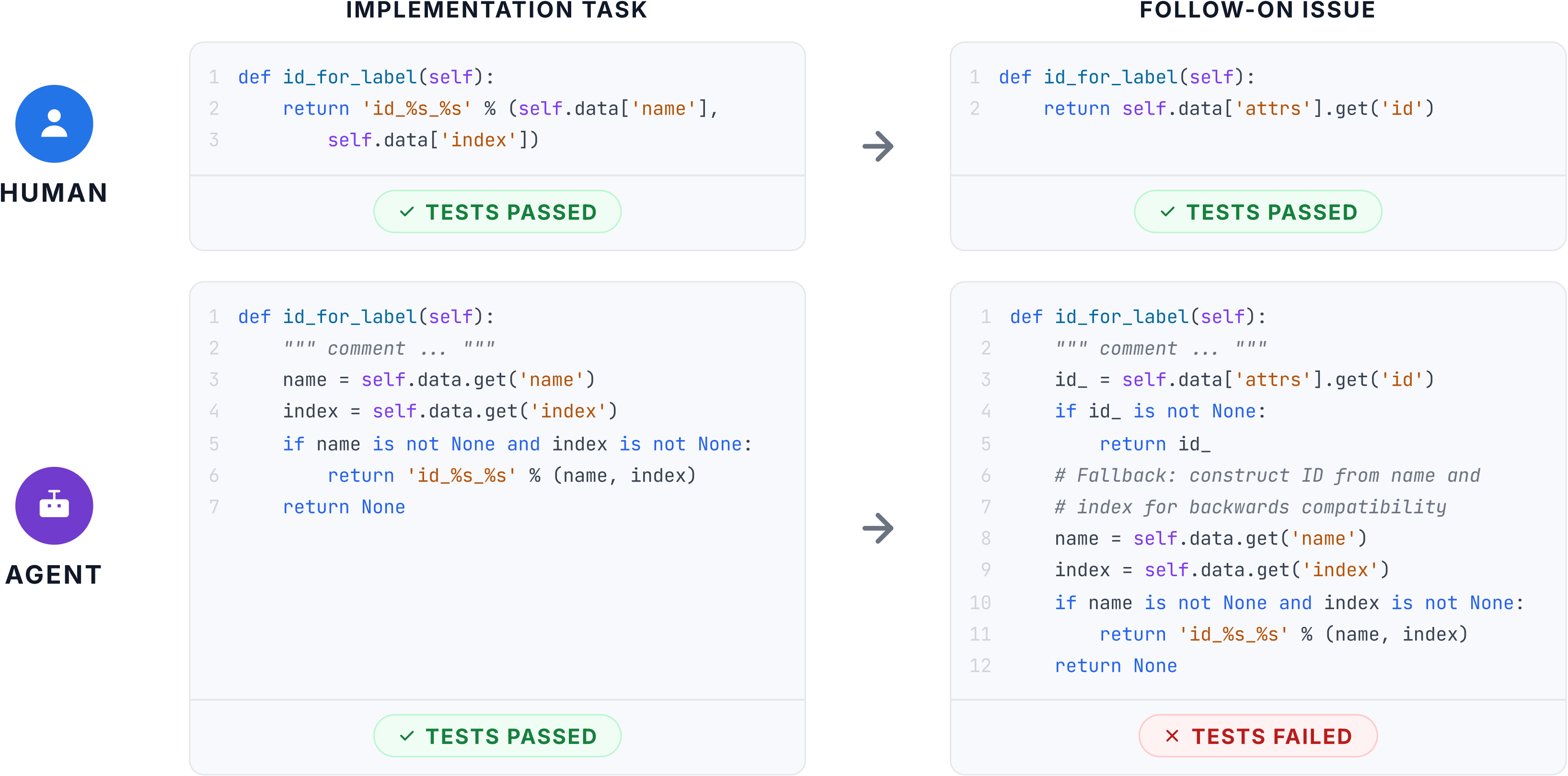}
    \caption{\textbf{Two pieces of code both being functionally correct does not mean they are equally maintainable.} An example instance where both the human and agent code pass the initial implementation task's tests, yet when an agent makes a subsequent change to both, the version built on human code passes while the version built on agent code fails.}
    \label{fig:agent_vs_human}
\end{figure*}

Coding agents are increasingly integrated into software engineering workflows, with promises of ``completing weeks of work in days'' \citep{openai_codex}. Model capabilities on coding tasks have also improved rapidly, with state-of-the-art agents solving a growing share of repository-level benchmarks \citep{openai2026swebenchverified}.
However, a patch that passes tests and is functionally correct may introduce structural complexity, brittle abstractions, or confusing implementation choices that only surface as failures when the code is extended or modified in downstream tasks (Figure \ref{fig:agent_vs_human}).

This dimension of code quality is captured by \textit{maintainability}, or how well the code can be modified and extended over time. Maintainability has been studied across many software development contexts, and static metrics have emerged to provide quantitative evaluations of maintainability. In particular, structural and size-based
metrics, such as cyclomatic complexity~\citep{cyclomaticComplexity},
Halstead volume~\citep{Halstead1977ElementsOS}, and cognitive
complexity~\citep{cognitiveComplexity}, have been used widely as proxies
for the maintenance cost of a project \citep{10.1109/ESEM.2009.5314233, aToolMaintainabilityMetric}. The adoption of coding agents raises questions about whether code authored by agents is as maintainable as human-authored code. Recent work has measured agent code maintainability using these static metrics~\citep{wang2025maintaincodermaintainablecodegeneration, slopcodebench}, but does not compare it with human code. Moreover, static metrics can often be confounded by codebase conventions and context~\citep{zhang2013does}.

To this end, we present \textbf{CodeThread}, a framework for controlled experiments to compare the maintainability of agent- and human-written code based on how well an agent can build on each. CodeThread transforms single-step, repository-level coding tasks (e.g., instances from SWE-Bench \cite{swebench}) into two-step tasks comprising an initial implementation task and a dependent downstream task. CodeThread then constructs comparable agent and human code on the initial task while holding the downstream task author and test-based evaluation fixed, allowing us to isolate how authorship of the initial task affects downstream agent performance.

We use CodeThread to evaluate four frontier models---Claude 4.5 Sonnet~\cite{sonnet45}, GPT-5~\cite{gpt5}, GLM 4.7~\cite{zeng2025glm, glm47}, and MiniMax 2.5~\cite{minimaxm25}---on four software engineering benchmarks spanning a range of programming languages and task types. We find that building on agent code leads to drops in downstream task resolve rates more often than building on human code, with gaps in resolve rate of up to 13.1\%. Using code instances from CodeThread, we identify sources of differences in downstream task performance on agent and human code, performing a logistic regression analysis on cases where the two conditions produce different outcomes. We find that traditional static maintainability metrics generally do not explain these differences. Instead, the clearest predictors of cases where the downstream task fails on agent code but succeeds on human code are subtle differences in input validation and error handling. Larger downstream edits and task difficulty also help characterize where the two conditions diverge. 

Our contributions are as follows:
\begin{enumerate}
    \item We introduce CodeThread, a  framework for studying the maintainability of agent code through two-step pull-request chains. CodeThread applies to any SWE benchmark, making it scalable across task types and languages, and provides controlled comparisons of agent to human code.
    
    \item Applying CodeThread to multiple frontier models and benchmarks, we show that agent code is overall less maintainable than human code, leading to worse downstream task performance. 
    
    \item We analyze the sources of maintainability differences between agent and human code, showing that traditional proxies such as structural complexity and verbosity do not fully capture these differences, which instead often arise from subtler behavioral effects.
    
\end{enumerate}
Together, CodeThread and our findings suggest that coding agents introduce maintainability costs that compound as their code is reused or extended in ways that current benchmarks and metrics fail to capture. We suggest directions for future work on the maintainability of agent code.
\section{Related Work}\label{section:related-works}

\begin{figure*}
    \centering
    \includegraphics[width=1.0\linewidth]{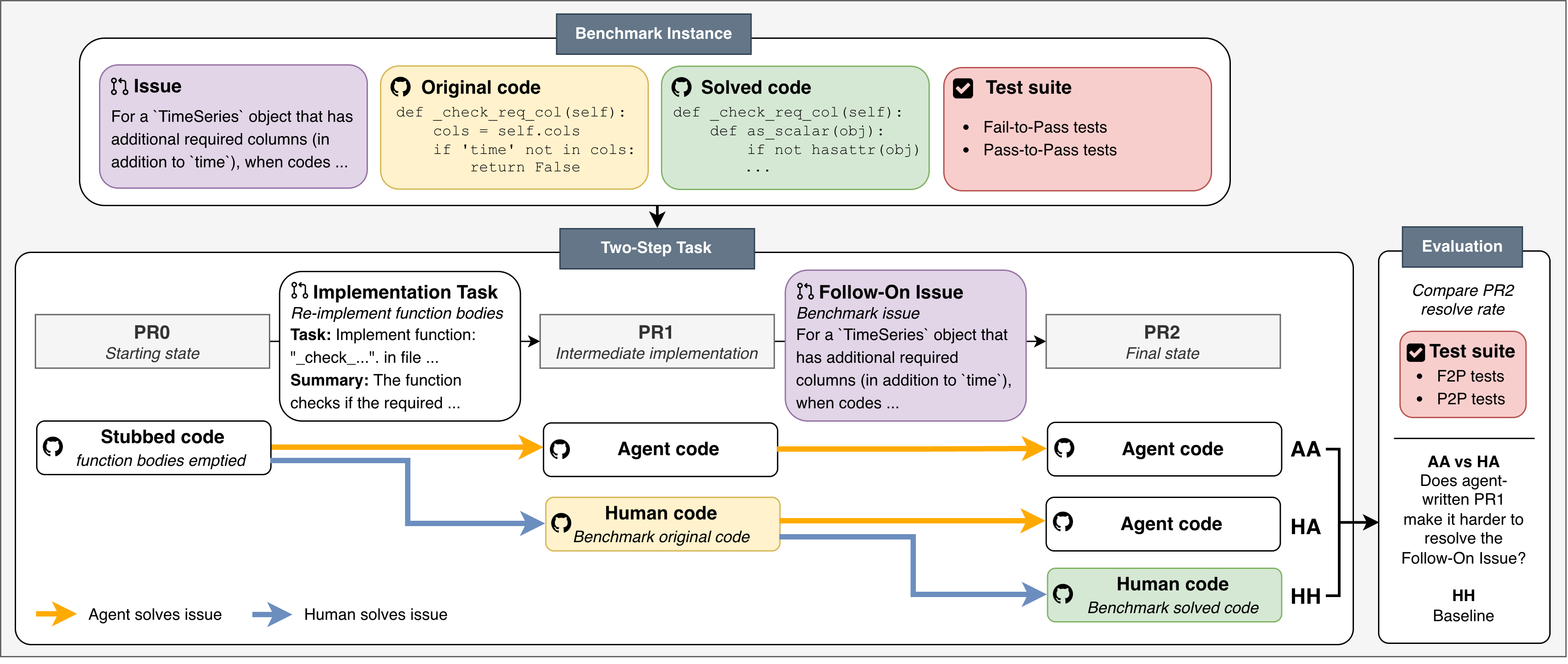}
    \caption{\textbf{CodeThread framework.} From the original benchmark instance, we construct a two-step task---an Implementation Task followed by a Follow-On Issue---producing three code states ($\text{PR}_0$, $\text{PR}_1$, and $\text{PR}_2$) and three conditions: AA (agent performs both steps), HA (human performs the Implementation Task, agent performs the Follow-On Issue on human code), and HH (human performs both steps). Comparing AA and HA isolates the effect of agent authorship on maintainability, holding the follow-on author fixed.}
    \label{fig:CodeThread-framework}
\end{figure*}

\paragraph{Evaluating Agents on Sequential Tasks.} 
Several recent benchmarks evaluate coding agents on sequences of dependent tasks rather than individual issues. 
CodeFlowBench decomposes Codeforces problems along their function-level dependency tree and requires the agent to implement each subfunction by reusing those it built in earlier turns~\citep{wang2026codeflowbenchmultiturniterativebenchmark}. 
SWE-Bench-CL orders GitHub issues chronologically within each repository to evaluate whether experience from earlier issues transfers to later ones \citep{joshi2025swebenchclcontinuallearningcoding}. 
SlopCodeBench measures behavioral drift as agents repeatedly modify a codebase across long-horizon iterative trajectories~\citep{slopcodebench}.
In contrast, our work grounds measurements in real GitHub pull requests rather than fully synthetic problems or hand-authored projects. Additionally, by varying the authorship of an intermediate patch while holding the downstream task fixed, we isolate the maintainability cost of agent-written code from the inherent difficulty of sequential development.

\paragraph{Maintainability of LLM-Generated Code.}
Large-scale GitHub analyses report rising code churn and duplication coinciding with AI-assistant adoption~\citep{harding2025copilot, cynthia2026beyond, huang2026codereuseinvestigatingcode, liu2026debtaiboomlargescale}, static analyses find that LLM-generated code exhibits more code smells than human-written code~\citep{paul2025investigatingsmellsllmgenerated}, and recent work shows that non-functional quality and design-constraint compliance degrade even when agent-generated patches pass tests \citep{sun2026quality, yu2026doespassratetell}. 
Prior work has also measured code-quality degradation in sequential tasks. 
MaintainCoder applies synthetic requirement changes to competitive-coding tasks and measures post-modification correctness alongside static metrics such as Cyclomatic Complexity and AST similarity~\citep{wang2025maintaincodermaintainablecodegeneration}. SWE-CI uses test pass rates over iterative edit trajectories as a maintainability proxy, treating later-stage pass rate as evidence of how well early decisions support subsequent maintenance~\citep{chen2026swecievaluatingagentcapabilities}. 

Our study instead measures maintainability directly through downstream issue-resolution performance. Rather than relying primarily on static proxies or synthetic requirement changes, we ask whether an agent can successfully complete a dependent follow-on task when the intermediate code it inherits is written by an agent versus a human, creating controlled experiments to measure the effect of authorship. The resulting code can also be analyzed using static metrics and failure-mode annotations, as performed in Section \ref{sec:analysis}.
\section{CodeThread}
We introduce CodeThread, a framework for measuring the downstream effects of building on agent code in comparison to human code (Figure~\ref{fig:CodeThread-framework}). CodeThread transforms standard single-issue software engineering benchmark instances into dependent two-step pull-request chains (Section~\ref{sec:CodeThread-create}), allowing for controlled comparisons of downstream outcomes across authorship conditions (Section~\ref{sec:CodeThread-authors}) with evaluation built in (Section~\ref{sec:CodeThread-evaluate}). 

CodeThread starts from an existing issue-resolution coding benchmark where instances consist of a written issue, a base commit, a human-authored ground-truth patch, and a test suite to determine whether the patch resolved the task. 
Example benchmarks that follow this format include SWE-Bench Verified~\citep{swebechverified}, SWE-Bench
Multilingual~\citep{swebenchmulti}, SWE-Bench
Pro~\citep{deng2025swebenchproaiagents}, and FeatBench~\citep{chen2025featbench}.

\subsection{Step 1: Create a Two-Step Task}
\label{sec:CodeThread-create}
From each benchmark instance, we construct a two-step chain consisting of two dependent tasks and three code states: the \textbf{Implementation Task} and \textbf{Follow-On Issue}, and code states \textbf{$\text{PR}_0$}, \textbf{$\text{PR}_1$}, and \textbf{$\text{PR}_2$}. To create the first state, {$\text{PR}_0$}, we remove the target function bodies while preserving their signatures, leaving a skeleton version of the benchmark’s initial code state. This provides a starting point from which different authors can implement the intended functionality, similar to the use of skeleton code in prototyping, where an initial code structure is later fleshed out with implementation details \cite{ryoo2008teaching, queiros2013codeskelgen}.

The Implementation Task (${\text{PR}_0} \rightarrow {\text{PR}_1}$) produces human and agent implementations of the same task. Using an LLM, we create an issue asking the author to restore the missing functionality in the skeletonized code. This issue is provided to the agent, which implements the requested functionality to produce the agent $\text{PR}_1$. We filter out $\text{PR}_1$ solutions that have already solved the benchmark issue by requiring them to pass all Pass-to-Pass tests, confirming that the original functionality is restored, while still failing all Fail-to-Pass tests, confirming that the downstream issue is not yet resolved. This ensures that the agent $\text{PR}_1$ reflects the benchmark's initial code state rather than a partial or complete solution to the Follow-On Issue; accordingly, the target methods are functionally equivalent across the two $\text{PR}_1$ versions. Although agents sometimes incidentally satisfy Fail-to-Pass tests while restoring functionality, the filter retains a large majority of instances. Appendix \ref{appendix:prompt_strategies} shows the prompt used to generate Implementation Task statements, as well as the prompt given to models when solving it.

The Follow-On Issue ($\text{PR}_1 \rightarrow \text{PR}_2$) is the original benchmark issue applied on top of the PR1 implementation. The problem statement and testing criteria for the Follow-On Issue are unchanged from the original benchmark, making it a downstream task whose difficulty may depend on the maintainability of the PR1 implementation. We evaluate whether the Follow-On Issue is resolved using the original benchmark test suite: a $\text{PR}_2$ solution is considered correct if it passes both the Pass-to-Pass tests, confirming that existing functionality is preserved, and the Fail-to-Pass tests, confirming that the original benchmark issue has been fixed. 

\subsection{Step 2: Set-up Authorship Scenarios}
\label{sec:CodeThread-authors}
Using the two-step issue resolution pipeline, we vary authorship on $\text{PR}_1$ and $\text{PR}_2$, creating a total of three comparable authorship scenarios: human $\text{PR}_1$ followed by human $\text{PR}_2$ (\textbf{HH}), human $\text{PR}_1$ followed by agent $\text{PR}_2$ (\textbf{HA}), and agent $\text{PR}_1$ followed by agent $\text{PR}_2$ (\textbf{AA}). 
The HH condition consists of the original developer's PRs provided by each benchmark, acting as the human baseline against which the other conditions are compared. Comparing AA to HA isolates the downstream effect of $\text{PR}_1$ authorship on $\text{PR}_2$ success, allowing us to directly evaluate whether agent code imposes a maintainability burden on subsequent agent work in comparison to human code. 

We focus on the maintainability costs that agents incur when building on prior code, comparing the outputs from the AA versus HA scenarios, as a growing fraction of developers use AI agents to extend existing codebases~\citep{li2025riseaiteammatessoftware}. The AH setting of a human performing the follow-on task on top of agent code would offer a complementary human-centric view (by comparing AH versus AA), but would require recruiting a substantial number of developers and is out of scope for this study. By keeping the $\text{PR}_2$ author fixed, CodeThread is scalable and can be applied to many existing issue-resolution benchmarks without additional data collection from humans.

\begin{table*}[t]
\centering
\small
\setlength{\tabcolsep}{4pt}
\begin{tabular}{l cc cc cc cc l}
\toprule
\textbf{Benchmark}
  & \multicolumn{2}{c}{\textbf{\# Instances}}
  & \multicolumn{2}{c}{\textbf{Filtered Subset}}
  & \multicolumn{2}{c}{\textbf{PR$_1$ (Avg.)}}
  & \multicolumn{2}{c}{\textbf{PR$_2$ (Avg.)}} 
  & \textbf{Task Types}\\
\cmidrule(lr){2-3} \cmidrule(lr){4-5} \cmidrule(lr){6-7} \cmidrule(lr){8-9}
 & Total & Filtered. & \#Repos & \#Lang & Files & Funcs & Files & Funcs \\
\midrule
SWE-Bench Verified     & 500 & 441 & 12 & 1 & 1.12 & 1.60 & 1.18 & 1.71 & BF  \\
SWE-Bench Multilingual & 300 & 171 & 34 & 7 & 1.29 & 1.90 & 1.84 & 3.28 & BF  \\
SWE-Bench Pro          & 731 & 641 & 11 & 4 & 2.43 & 5.51 & 5.22 & 10.70 & BF, FI \& RF \\
FeatBench              & 156 & 124 & 22 & 1 & 1.52 & 2.54 & 2.35 & 4.03 & FI  \\
\midrule
\textbf{Total} & \textbf{1{,}687} & \textbf{1{,}377} & 78 & 10 & 1.79 & 3.54 & 3.25 & 6.30 \\
\bottomrule 
\end{tabular}
\caption{\textbf{Benchmarks used in our CodeThread evaluation.} We list each benchmark's full and filtered instance counts, retaining only instances requiring function-level edits. For the filtered subset, we report repository and language coverage (\#Repos, \#Lang), the mean files and functions modified per PR stage ($\text{PR}_1$, $\text{PR}_2$), and a breakdown of task types: bug fixes (BF), feature implementation (FI), and refactoring (RF).}
\label{tab:benchmark_comparison}
\end{table*}

\subsection{Step 3: Evaluate agent vs human code}
\label{sec:CodeThread-evaluate}

We assess the final HA and AA outputs, $\text{PR}_2$, using resolve rate: the percentage of instances in which the code successfully resolves the original benchmark issue after both editing steps. Differences in resolve rate indicate whether the authorship of $\text{PR}_1$ affects downstream issue-resolution success, providing a task-based measure of maintainability. The resulting code samples can also be compared along other dimensions of code quality since the task setup is controlled and the downstream issue is held fixed across authorship conditions. We show this in Section \ref{sec:analysis}.

\section{CodeThread on SWE Problems}
\subsection{Experimental Set-up}

\paragraph{Task Types.} We source a total of 1,377 instances from SWE-Bench Verified~\citep{swebechverified}, SWE-Bench Multilingual~\citep{swebenchmulti}, SWE-Bench Pro~\citep{deng2025swebenchproaiagents}, and FeatBench~\citep{chen2025featbench}. 
The four benchmarks were selected to cover a range of software engineering tasks: bug fixing (BF), feature implementation (FI), and refactoring (RF), as shown in Table \ref{tab:benchmark_comparison}.
Following \citet{xu2025swecompassunifiedevaluationagentic}, we treat SWE-Bench Verified and SWE-Bench Multilingual as bug fix benchmarks.  We use the native labels of each SWE-Bench Pro instance, and treat FeatBench as a feature-implementation benchmark.

\paragraph{Models and Agent Configuration.} In order to cover a breadth of model types, we evaluate four frontier models: Claude 4.5 Sonnet, GPT-5, GLM 4.7 and MiniMax M2.5, two proprietary frontier models and two publicly available open-weight models. We use SWE-Agent \cite{yang2024sweagent} as the scaffolding framework for all models. We follow the standard evaluation protocol of the underlying benchmarks and run each pair (model, authorship condition) once per instance in order to maintain tractable computational costs while providing coverage across all benchmark categories. We provide further details in Appendix \ref{appendix:executor-details}.

\subsection{Results}

\textbf{Overall findings.} Table \ref{tab:pr2_ha_vs_aa} shows the $\text{PR}_2$ resolve rates under HA and AA conditions. The number of instances that pass the $\text{PR}_1$ filtering differs by model, so we compare models using the subset of instances shared across all models within each benchmark. Across these instances, we find that agents more often perform worse when building on agent code than on human code: \textbf{AA generally underperforms HA, with drops in downstream resolve rate of up to 13.1\%.} Full scores for each model and benchmark are provided in Appendix \ref{appendix:resolve_rate_full_data}; we observe a similar pattern in both the shared-instance analysis and the full per-model benchmark results. The largest drops occur for GLM 4.7 on SWE-Bench Pro, which contains longer, multi-file edits drawn from professional codebases, and with GPT-5 on FeatBench, a set of feature implementation tasks. We trace the source of the gap between HA and AA in Section~\ref{sec:analysis}. 

We also observe four cases where AA matches or exceeds HA performance. These should be considered in light of benchmark-specific baseline performance and sample size. GPT-5 shows small positive gaps on SWE-Bench Multilingual (+3.1) and SWE-Bench Pro (+0.8), but these coincide with lower HA baseline performance on the benchmarks, suggesting that the gap may reflect a weaker ceiling rather than greater robustness to agent code. The clearest positive result, Claude 4.5 Sonnet on FeatBench (+6.2), occurs on our smallest overlap set of 16 instances, where a small number of cases can substantially shift the observed gap.

\paragraph{Per-model patterns.} 
Table~\ref{tab:pr2_ha_vs_aa} shows the HA-vs-AA gap across models. GLM shows the largest and most consistent effect of $\text{PR}_1$ authorship, with AA underperforming HA on all benchmarks and dropping by 13.1\% on SWE-Bench Pro. Claude and MiniMax also drop by around $3$-$8$\% across SWE-Bench variants, though FeatBench shows mixed results. GPT-5 has the smallest gaps on SWE-Bench variants, however, we see a 12.5\% drop on FeatBench, suggesting that maintainability costs may be more visible on feature implementation tasks. 

\begin{table*}[t]
      \centering
      \definecolor{negcolor}{rgb}{0.70,0.13,0.13}
      \definecolor{poscolor}{rgb}{0.0,0.6,0.3}
      \small
      \setlength{\tabcolsep}{4.5pt}

      \begin{tabular}{l ccc ccc ccc ccc}
          \toprule
           & \multicolumn{12}{c}{\textbf{Resolve Rate \% ($\uparrow$)}} \\
          \cmidrule(lr){2-13}
           & \multicolumn{3}{c}{\textbf{Claude}} & \multicolumn{3}{c}{\textbf{GPT}} & \multicolumn{3}{c}{\textbf{GLM}} & \multicolumn{3}{c}{\textbf{MiniMax}} \\
          \cmidrule(lr){2-4} \cmidrule(lr){5-7} \cmidrule(lr){8-10} \cmidrule(lr){11-13}
          \textbf{Benchmark} & HA & AA & $\Delta$ & HA & AA & $\Delta$ & HA & AA & $\Delta$ & HA & AA & $\Delta$ \\
          \midrule
          SWE-Bench Verified
              & \textbf{71.7} & 66.7 & \cellcolor{negcolor!21}\textcolor{negcolor}{$-$5.1}
              & \textbf{62.6} & 61.6 & \cellcolor{negcolor!4}\textcolor{negcolor}{$-$1.0}
              & \textbf{65.7} & 62.6 & \cellcolor{negcolor!13}\textcolor{negcolor}{$-$3.0}
              & \textbf{74.7} & 68.7 & \cellcolor{negcolor!26}\textcolor{negcolor}{$-$6.1} \\
          SWE-Bench Multilingual
              & \textbf{53.1} & 50.0 & \cellcolor{negcolor!13}\textcolor{negcolor}{$-$3.1}
              & 46.9 & \textbf{50.0} & \cellcolor{poscolor!13}\textcolor{poscolor}{$+$3.1}
              & \textbf{62.5} & 59.4 & \cellcolor{negcolor!13}\textcolor{negcolor}{$-$3.1}
              & \textbf{65.6} & 62.5 & \cellcolor{negcolor!13}\textcolor{negcolor}{$-$3.1} \\
          SWE-Bench Pro
              & \textbf{46.3} & 38.6 & \cellcolor{negcolor!32}\textcolor{negcolor}{$-$7.7}
              & 39.0 & \textbf{39.8} & \cellcolor{poscolor!3}\textcolor{poscolor}{$+$0.8}
              & \textbf{45.2} & 32.0 & \cellcolor{negcolor!55}\textcolor{negcolor}{$-$13.1}
              & \textbf{43.6} & 36.7 & \cellcolor{negcolor!29}\textcolor{negcolor}{$-$6.9} \\
          FeatBench
              & 50.0 & \textbf{56.2} & \cellcolor{poscolor!26}\textcolor{poscolor}{$+$6.2}
              & \textbf{56.2} & 43.8 & \cellcolor{negcolor!52}\textcolor{negcolor}{$-$12.5}
              & \textbf{62.5} & 56.2 & \cellcolor{negcolor!26}\textcolor{negcolor}{$-$6.2}
              & 56.2 & 56.2 & 0.0 \\
          \bottomrule
      \end{tabular}
      \caption{\textbf{Comparing HA and AA PR$_2$ resolve rate.} We observe a consistent lower resolution rate under the AA condition than HA across nearly all model--benchmark pairs. Bolded values indicate the higher resolve rate within
  each HA/AA pair.}
      \label{tab:pr2_ha_vs_aa}
  \end{table*}

\begin{table*}[t]
      \centering
      \definecolor{negcolor}{rgb}{0.70,0.13,0.13}
      \definecolor{poscolor}{rgb}{0.0,0.6,0.3}
      \small
      \setlength{\tabcolsep}{4pt}

      \begin{tabular}{l c ccc ccc ccc ccc c}
          \toprule
           & & \multicolumn{12}{c}{\textbf{Resolve Rate \% ($\uparrow$)}} & \\
          \cmidrule(lr){3-14}
           & & \multicolumn{3}{c}{\textbf{Claude}} & \multicolumn{3}{c}{\textbf{GPT}} & \multicolumn{3}{c}{\textbf{GLM}} & \multicolumn{3}{c}{\textbf{MiniMax}} & \\
          \cmidrule(lr){3-5} \cmidrule(lr){6-8} \cmidrule(lr){9-11} \cmidrule(lr){12-14}
          \textbf{Task} & \textbf{N} & HA & AA & $\Delta$ & HA & AA & $\Delta$ & HA & AA & $\Delta$ & HA & AA & $\Delta$ & \textbf{Avg. $\Delta$} \\
          \midrule
          BF & 210
              & \textbf{59.05} & 51.43 & \cellcolor{negcolor!23}\textcolor{negcolor}{$-$7.62}
              & 49.52 & \textbf{50.00} & \cellcolor{poscolor!1}\textcolor{poscolor}{$+$0.48}
              & \textbf{56.19} & 52.38 & \cellcolor{negcolor!12}\textcolor{negcolor}{$-$3.81}
              & \textbf{61.43} & 56.19 & \cellcolor{negcolor!16}\textcolor{negcolor}{$-$5.24}
              & \cellcolor{negcolor!12}\textcolor{negcolor}{$-$4.05} \\
          FI & 104
              & \textbf{44.23} & 41.35 & \cellcolor{negcolor!9}\textcolor{negcolor}{$-$2.88}
              & \textbf{41.35} & 38.46 & \cellcolor{negcolor!9}\textcolor{negcolor}{$-$2.88}
              & \textbf{46.15} & 30.77 & \cellcolor{negcolor!47}\textcolor{negcolor}{$-$15.38}
              & \textbf{42.31} & 38.46 & \cellcolor{negcolor!12}\textcolor{negcolor}{$-$3.85}
              & \cellcolor{negcolor!19}\textcolor{negcolor}{$-$6.25} \\
          RF & 70
              & \textbf{57.14} & 50.00 & \cellcolor{negcolor!22}\textcolor{negcolor}{$-$7.14}
              & 48.57 & \textbf{51.43} & \cellcolor{poscolor!9}\textcolor{poscolor}{$+$2.86}
              & \textbf{55.71} & 40.00 & \cellcolor{negcolor!48}\textcolor{negcolor}{$-$15.71}
              & \textbf{54.29} & 41.43 & \cellcolor{negcolor!39}\textcolor{negcolor}{$-$12.86}
              & \cellcolor{negcolor!25}\textcolor{negcolor}{$-$8.21} \\
          FI + RF & 22
              & \textbf{27.27} & 22.73 & \cellcolor{negcolor!14}\textcolor{negcolor}{$-$4.55}
              & 27.27 & 27.27 & 0.00
              & \textbf{31.82} & 13.64 & \cellcolor{negcolor!55}\textcolor{negcolor}{$-$18.18}
              & \textbf{27.27} & 22.73 & \cellcolor{negcolor!14}\textcolor{negcolor}{$-$4.55}
              & \cellcolor{negcolor!21}\textcolor{negcolor}{$-$6.82} \\
          \bottomrule
      \end{tabular}
      \caption{\textbf{Task-type stratified resolved rate results by model on overlapping instances. }The dataset is stratified by task type into bug fixes (BF), feature implementation (FI), refactoring (RF), and feature implementation \&
  refactoring (FI + RF). PR$_2$ resolve rates under HA and AA; $\Delta$ denotes the percentage-point change from HA to AA within each task type and model. Avg. $\Delta$ denotes the average \% change across models for each task type.}
      \label{tab:task-type-resolved-rate-model-results}
  \end{table*}

\paragraph{Stratifying by task type.} Table~\ref{tab:task-type-resolved-rate-model-results} shows that refactoring tasks have the largest average drop across models of 8.21\%. 
Feature implementation and combined FI+RF tasks also show notable average drops of 6.25 and 6.82\%, respectively.
Bug-fix tasks show the smallest average drop, with an average 4.05\%. 
These observations are consistent with prior work, which show that refactoring and multi-file edits are harder for coding agents than localized fixes \cite{xu2025swecompassunifiedevaluationagentic, gautam2025refactorbenchevaluatingstatefulreasoning}. 

\section{Why does AA underperform HA?}\label{sec:analysis}
We observe that agents building on agent code generally perform worse than on human code. To investigate the drivers of this gap, we focus on discordant instances---i.e., those where HA and AA yield different outcomes on the task.
Across all four models and benchmarks, there are 454 discordant instances. 64.3\% of these are HA wins (HA passes while AA fails), while 35.7\% are AA wins (AA passes while HA fails). After excluding instances with incomplete features, our analysis includes 405 instances (64.7\% HA wins, 35.3\% AA wins). We extract features from $\text{PR}_1$ and $\text{PR}_2$, detailed in Section \ref{sec:lr-features}, and fit a logistic regression model to identify which features most strongly predict whether HA or AA won in a discordant instance.

\subsection{Features}\label{sec:lr-features}
We use four feature categories to identify the root causes of the gap: static maintainability metrics to test traditional code-quality signals, patch localization features to compare the files and functions edited, behavioral drift labels to capture differences missed by static metrics, and an instance difficulty score to control for task-level variation. We describe each below and provide implementation details in Appendix~\ref{appendix:lr-feature-details}.

\paragraph{Static maintainability metrics on $\text{PR}_1$ and $\text{PR}_2$.}
Our first feature set comprises four static proxies for maintenance cost based on software engineering literature~\citep{10.1109/ESEM.2009.5314233, aToolMaintainabilityMetric}, spanning two dimensions of code quality. 
The first dimension is \textit{structural complexity}, for which we use Cyclomatic Complexity (CC)~\citep{cyclomaticComplexity}, which counts linearly independent paths through a function, and Cognitive Complexity (CogC)~\citep{cognitiveComplexity}, which penalizes nested control structures. 
The second dimension is \textit{verbosity}, for which we use Halstead Volume (HV)~\citep{Halstead1977ElementsOS}, based on operator and operand counts, and Logical Lines of Code (LLOC), which counts executable statements. For each metric, we compute two features per discordant instance: the AA--HA delta at $\text{PR}_1$ and $\text{PR}_2$. 

\definecolor{HAwinText}{rgb}{0.22, 0.55, 0.60}
\definecolor{AAwinText}{rgb}{0.45, 0.42, 0.70}

\begin{figure*}[t]
    \centering
    \includegraphics[width=\linewidth]{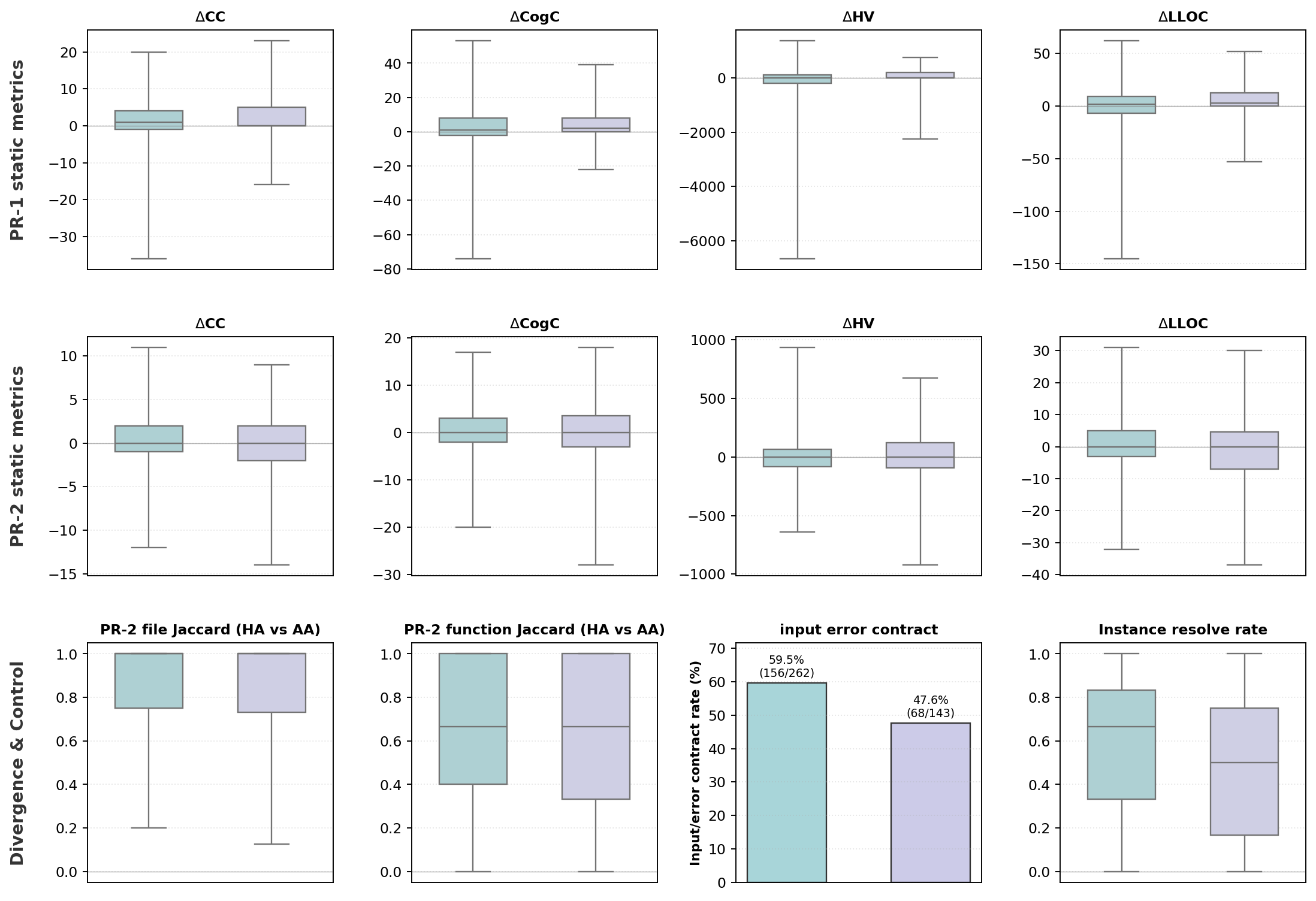}
    \caption{\textbf{HA vs AA-only wins are differentiated by behavioral drift, instance difficulty, and downstream code-size change.} Panels compare instances resolved only in HA (\textcolor{HAwinText}{teal}) versus in AA (\textcolor{AAwinText}{lavender}); whiskers show the 5th/95th percentiles. Most static maintainability metrics have similar distributions, except $\Delta$LLOC at $\text{PR}_2$; input error contract and instance resolve rate also separate the two groups.}
    \label{fig:disagreement-distributions}
\end{figure*}

\paragraph{Patch localization features on $\text{PR}_2$.}
Our second feature set measures files and functions that the agent edits. 
Prior work has shown that localization is one of the leading failure modes for coding agents~\citep{he2026sweadeptllmbasedagenticframework, liu2025empiricalstudyfailuresautomated}. We focus on $\text{PR}_2$ because the $\text{PR}_1$ prompt explicitly lists the files and functions each chain must edit. The $\text{PR}_2$ problem statement does not, and the agent can freely choose which files and functions to edit.
We measure this with two Jaccard overlaps: one over the set of edited files, and one over the set of edited functions. Higher values indicate that the HA and AA patches modify more similar parts of the codebase. 

\paragraph{Behavioral drift labels on $\text{PR}_1$.}
The third feature set captures differences between agent and human $\text{PR}_1$s not reflected in aggregate static metrics. We develop a ten-category taxonomy for forms of behavioral drift, detailed in Appendix \ref{appendix:pr1-fidelity-judge}, and use Deepseek-V4 Pro~\citep{deepseekv4} as an LLM-judge to label instances by category. 
We are particularly interested in cases where the agent's $\text{PR}_1$ alters input-validation or error-handling behavior, which we define as \textit{input/error contract} (IEC). From our observations, differences in exception types or swapped defaults can silently creep into the codebase. Agents may add an extra input-gating check that raises \texttt{ValueError} or use \texttt{getattr} with \texttt{None} as the default, while the human $\text{PR}_1$ would raise an error. In both cases, neither static metrics nor patch location features would distinguish the agent from the human code. Appendix~\ref{sec:iec_examples} shows five such cases contrasting the human and agent implementations of $\text{PR}_1$ and $\text{PR}_2$.
As such, we look at whether or not the behavioral drift was labeled as one of three IEC categories---Input Gating, Silent Fallback, and Error Handling---setting the feature to 1 if so, and 0 otherwise. To verify LLM-judge accuracy, we manually check a subset of 50 LLM-judge labels (Appendix \ref{appendix:human-audit}).

\paragraph{Instance difficulty control.}
We add a difficulty control to ensure that HA is not winning simply because it is evaluated on easier tasks. We use a leave-one-out (LOO) instance difficulty score, defined as the mean resolution rate of the other three models on the same instance. 

\subsection{Modeling}

Given these twelve features, together with benchmark and model fixed effects, we fit a logistic regression (LR) to identify which features push the outcome toward HA or AA winning.
We define $Y_i = 1$ when HA resolves and AA fails on instance $i$ and $Y_i=0$ otherwise. 
The model is defined as:
\begin{equation*}
  \begin{aligned}
  \operatorname{logit}\,\Pr(Y_i = 1 \mid X_i)
  &= \beta_0 \,+\, \beta_{\text{IEC}}\,\text{IEC}_i +\, \boldsymbol{\beta}_J^{\!\top}\,\mathbf{J}_i \\
  &\quad +\, \boldsymbol{\beta}_1^{\!\top}\,\boldsymbol{\Delta}^{\text{PR}_1}_i
          \,+\, \boldsymbol{\beta}_2^{\!\top}\,\boldsymbol{\Delta}^{\text{PR}_2}_i \\
  &\quad \,+\, \boldsymbol{\beta}_D^{\!\top}\,\mathbf{D}_i 
          \,+\, \beta_{\text{LOO}}\,\text{LOO}_i
  \end{aligned}
  \label{eq:disagreement-lr}
\end{equation*}

Static maintainability metrics are denoted with $\boldsymbol{\Delta}^{\text{PR}_1}_i$ and $\boldsymbol{\Delta}^{\text{PR}_2}_i$ for each of four AA$-$HA metric deltas (CC, CogC, HV, LLOC) at $\text{PR}_1$ and $\text{PR}_2$. For fault localization, the HA-vs-AA file and function Jaccards on the $\text{PR}_2$ patches are denoted by $\mathbf{J}_i = (J^{\text{file}}_i, J^{\text{func}}_i)$. $\text{PR}_1$ behavioral drift is denoted by $\text{IEC}_i \in \{0,1\}$, an input/error-contract indicator on the agent's $\text{PR}_1$. $\text{LOO}_i$ denotes the leave-one-out instance difficulty score, and $\mathbf{D}_i$ is a fixed-effect for benchmark and model. We interpret coefficients as Odds Ratios (ORs), each of which captures how a feature shifts the outcome: a one-unit increase in feature $k$ multiplies the odds of HA winning by $\text{OR}_k$. Coefficients with OR $> 1$ therefore tilt the outcome toward HA, and OR $< 1$ toward AA. In Figure \ref{fig:disagreement-distributions}, we similarly observe the effects of each feature on the direction of discordance by visualizing the distribution of each feature across HA wins and AA wins.

\subsection{Findings}

\begin{table}[t]
    \centering
    \small
    \setlength{\tabcolsep}{10pt}
    \begin{tabular}{l r r}
        \toprule
        \textbf{Feature} & \textbf{OR (95\% CI)} & $\boldsymbol{p}$ \\
        \midrule
        $\Delta\text{CC}_{\text{PR}_1}$       & $0.92$ ($0.53$, $1.61$) & $0.779$ \\
        $\Delta\text{CogC}_{\text{PR}_1}$     & $1.00$ ($0.67$, $1.48$) & $0.981$ \\
        $\Delta\text{HV}_{\text{PR}_1}$       & $1.03$ ($0.69$, $1.53$) & $0.885$ \\
        $\Delta\text{LLOC}_{\text{PR}_1}$     & $0.92$ ($0.53$, $1.61$) & $0.780$ \\
        \midrule
        $\Delta\text{CC}_{\text{PR}_2}$       & $0.71$ ($0.38$, $1.31$) & $0.271$ \\
        $\Delta\text{CogC}_{\text{PR}_2}$     & $0.97$ ($0.69$, $1.37$) & $0.866$ \\
        $\Delta\text{HV}_{\text{PR}_2}$       & $0.93$ ($0.69$, $1.25$) & $0.615$ \\
        $\Delta\text{LLOC}_{\text{PR}_2}$     & $\mathbf{1.88}$ ($1.02$, $3.46$) & $\mathbf{0.042}$ \\
        \midrule
        File Jaccard         & $0.77$ ($0.58$, $1.01$) & $0.061$ \\
        Func. Jaccard        & $1.22$ ($0.93$, $1.60$) & $0.144$ \\
        \midrule
        IEC                  & $\mathbf{1.83}$ ($1.15$, $2.92$) & $\mathbf{0.011}$ \\
        \midrule
        Inst.\ resolve rate  & $\mathbf{1.42}$ ($1.12$, $1.78$) & $\mathbf{0.003}$ \\
        \bottomrule
    \end{tabular}
    \caption{\textbf{Effect of $\text{PR}_1$ and $\text{PR}_2$ features on HA-to-AA Odds Ratio (OR).} The three significant predictors of whether HA or AA will resolve in the discordant pair are: $\Delta\text{LLOC}_{\text{PR}_2}$, IEC, and instance difficulty. All three tilt the outcome toward HA: IEC drift in agent $\text{PR}_1$, AA over-editing at $\text{PR}_2$ relative to HA, and easier instances each raise the odds that only HA resolves.}
    \label{tab:disagreement-lr}
\end{table}

\paragraph{Which features predict HA and AA disagreement?} Table~\ref{tab:disagreement-lr} reports the fitted logistic regression on the discordant instances. Three features emerge as statistically significant predictors of the discordance direction, with all three tilting the outcome toward HA winning: IEC, $\Delta\text{LLOC}_{\text{PR}_2}$, and Instance resolve rate. 
When agent $\text{PR}_1$ alters input-validation or error-handling behavior, the odds of HA winning are $1.83\times$ that of AA, and when AA's $\text{PR}_2$ adds more lines than HA's, HA is $1.88\times$ more likely to win over AA. 
The difficulty control feature signifies that instances where HA wins tend to be easier. 
The model recovers a modest but reliable signal (McFadden $R^2 = 0.069$, LR $\chi^2(18) = 36.1$, $p = 0.007$) suggesting that these features are informative signals of downstream success, but do not fully explain or predict it. 

Figure~\ref{fig:disagreement-distributions} corroborates these results as static maintainability metrics show near-identical distributions on HA wins and AA wins, while IEC drift, $\Delta\text{LLOC}_{\text{PR}_2}$, and instance resolve rate show the clearest visual separation between the two groups. To verify the LLM-judge labels and assess robustness to the choice of judge, we re-ran the $\text{PR}_1$ behavioral drift judge with Kimi K2.6~\citep{kimik26} and re-fit the same model. This yielded the same three significant features with comparable odds ratios (IEC: OR $=1.70$, $p=0.039$; see Appendix~\ref{appendix:judge-robustness} for further details).

\paragraph{Does $\text{PR}_1$ drift cause the unresolved outcome in $\text{PR}_2$?}
The IEC coefficient in Table~\ref{tab:disagreement-lr} establishes the population-level signal, but population-level evidence does not tell us whether the drift was actually the cause of failing to resolve a given instance. We further investigate the cause of failures in the AA scenario, focusing on IEC for per-instance attribution as the only significant predictor in our model that identifies a specific, code-level mechanism. $\Delta\text{LLOC}_{\text{PR}_2}$ is a summary statistic and instance difficulty is a control variable, neither of which corresponds to a verifiable property of the agent's $\text{PR}_1$. 

We consider two scenarios that could decouple IEC drift from the downstream failure. First, the agent may rewrite the function containing the $\text{PR}_1$ input/error-contract divergence in $\text{PR}_2$, removing the drift before evaluation. Second, the drift may persist into $\text{PR}_2$, but the failing test may be evaluating a different part of the code. We use an LLM-judge (DeepSeek-V4 Pro) to attribute whether the failing test falls under either of these scenarios. The LLM-judge is given the $\text{PR}_1$ behavioral drift label with the generated rationale, the AA $\text{PR}_2$ patch, and the AA test outcome. It judges whether the divergence persisted into $\text{PR}_2$, and whether the failing test can be traced back to the divergence.

Table~\ref{tab:pr1-drift-attribution} reports the breakdown. The first scenario is rare: $\text{PR}_2$ rewrites the divergent function on only 6.9\% of instances, while the drift is unchanged in 85.9\% of instances. On 20.6\% of instances, the drift is judged to both persist and directly relate to the failing test case. The remaining 74.4\% fail for reasons we have not isolated, and we leave this analysis to future work. The high persistence rate also suggests that on long-horizon tasks (i.e., chains longer than our two-PR setup), this drift could compound across PRs and degrade downstream resolve rates further. Our setup cannot test this directly, and future work should assess the effects of these drifts over long-horizon tasks. The full prompt and output schema are provided in Appendix~\ref{appendix:pr2-pr1-judge}.

\section{Conclusion}
Across four benchmarks and four frontier models, building on agent code more often lowers downstream resolve rates than building on human code, with effects varying by model and task type. These differences are not well explained by static maintainability metrics alone; the clearest code-level signal is subtle behavioral drift in agent code, while larger downstream edits and task difficulty also indicate where the conditions diverge. Thus, building on agent code often introduces maintainability costs, which appear even on two-step chains and are likely to compound over many edits of a software project. As agent contributions become a larger share of working codebases, evaluation must shift from isolated task completion toward long-term maintainability.

\section*{Limitations}

We present limitations of our study. First, we consider only two-step pull request chains (\textit{PR$_1$} $\rightarrow$ \textit{PR$_2$}). Although this setup is sufficient to study whether the quality of an intermediate patch affects a downstream modification, it does not capture longer-horizon development trajectories, where technical debt may compound over multiple successive edits. Future work should explore constructing longer PR chains with function-level overlap across successive tasks.

Second, we filter $\text{PR}_1$ by instances where the agent $\text{PR}_1$ did not already address the Follow-On Issue. This filtering is necessary for the two-step chain to be well-defined, maintaining the separation between the two tasks so that the effect of authorship can be evaluated in isolation from functionality. However, it may introduce selection bias, since restricting the analyzed instances to where the agent did not solve the Follow-On Issue at $\text{PR}_1$ may not represent the agent's natural behavior across the full benchmark. Nonetheless, the filter retains a large majority of instances, and we verify that cases where the model satisfies Fail-to-Pass tests in the Implementation Task are often incidental. 

Third, as discussed in Section \ref{sec:CodeThread-authors}, our work does not study the effect of human-authored $\text{PR}_2$ on top of agent-authored $\text{PR}_1$ (AH) setting. Doing so would allow for a comparison of how difficult it is for human developers to build on agent code versus human code, and provide a more human-centric analysis of code maintainability. However, constructing this condition at scale would require recruiting developers to author $\text{PR}_2$ across all instances, which is infeasible in our current setting and difficult to control due to factors such as familiarity with the codebase. Future work could examine the AH setting on a smaller curated subset.  

\section*{Ethical Considerations}

This work aims to support more reliable evaluation and deployment of coding agents by studying the maintainability of code produced by these systems. Our findings show that agent code can affect downstream maintenance, underscoring the need for benchmarks, review practices, and safeguards that evaluate coding agents beyond immediate task completion and support informed decisions about when and how to rely on agent code. These findings should not be interpreted as evidence against the use of coding agents broadly, but as a call for evaluations that better reflect the long-term effects on software development.

\section*{Acknowledgments}
We thank Carlos Jimenez, Nicholas Lourie, Varun Yerram, Shashwat Singh, Rico Angell, and Yueh-Han Chen for insightful feedback and discussion. This work was supported by NSF IIS-2340345, NSF CAREER Award 2145542, and Coefficient Giving (formerly Open Philanthropy), and in part by the NYU IT High Performance Computing resources, services, and staff expertise. 

\bibliography{custom}

\clearpage

\appendix
\renewcommand{\thefigure}{A\arabic{figure}}
\renewcommand{\thetable}{A\arabic{table}}

\setcounter{figure}{0}
\setcounter{table}{0}

\section{Synthetic Problem Statement Construction for $\text{PR}_1$}\label{appendix:prompt_strategies}

\begin{figure*}[h]
     \centering
      \includegraphics[width=1.0\textwidth]{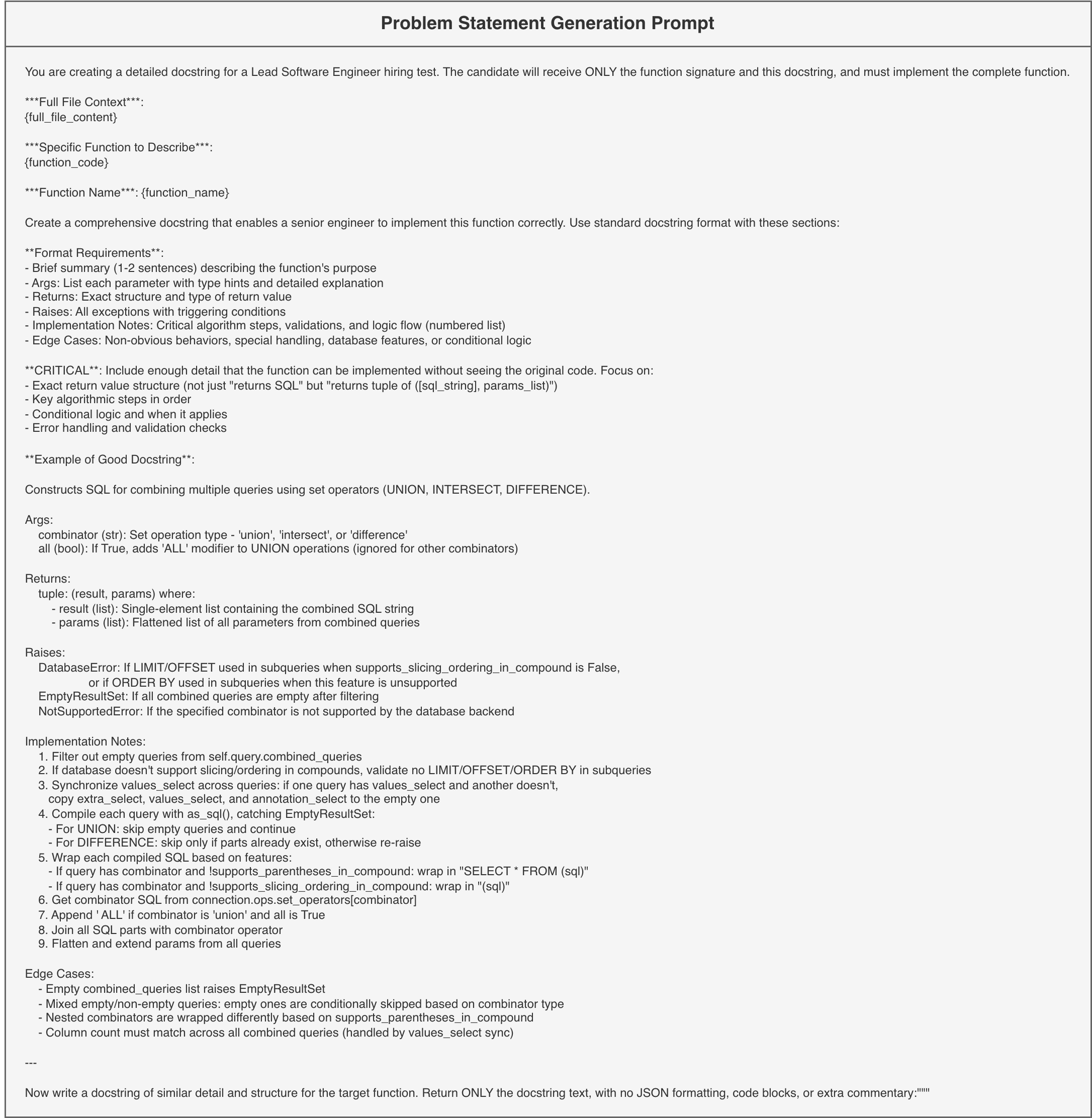} 
    \caption{\textbf{Prompt used to generate synthetic $\text{PR}_1$ problem statements.} Given the full file context, target function code, and function name, the model is instructed to produce a standalone docstring that enables implementation without revealing the original function body. The prompt specifies the required sections: summary, arguments, returns, exceptions, implementation notes, and edge cases. }
    \label{fig:pr_1_ps_prompt}
\end{figure*}

\begin{figure*}[h]
     \centering
      \includegraphics[width=1.0\textwidth]{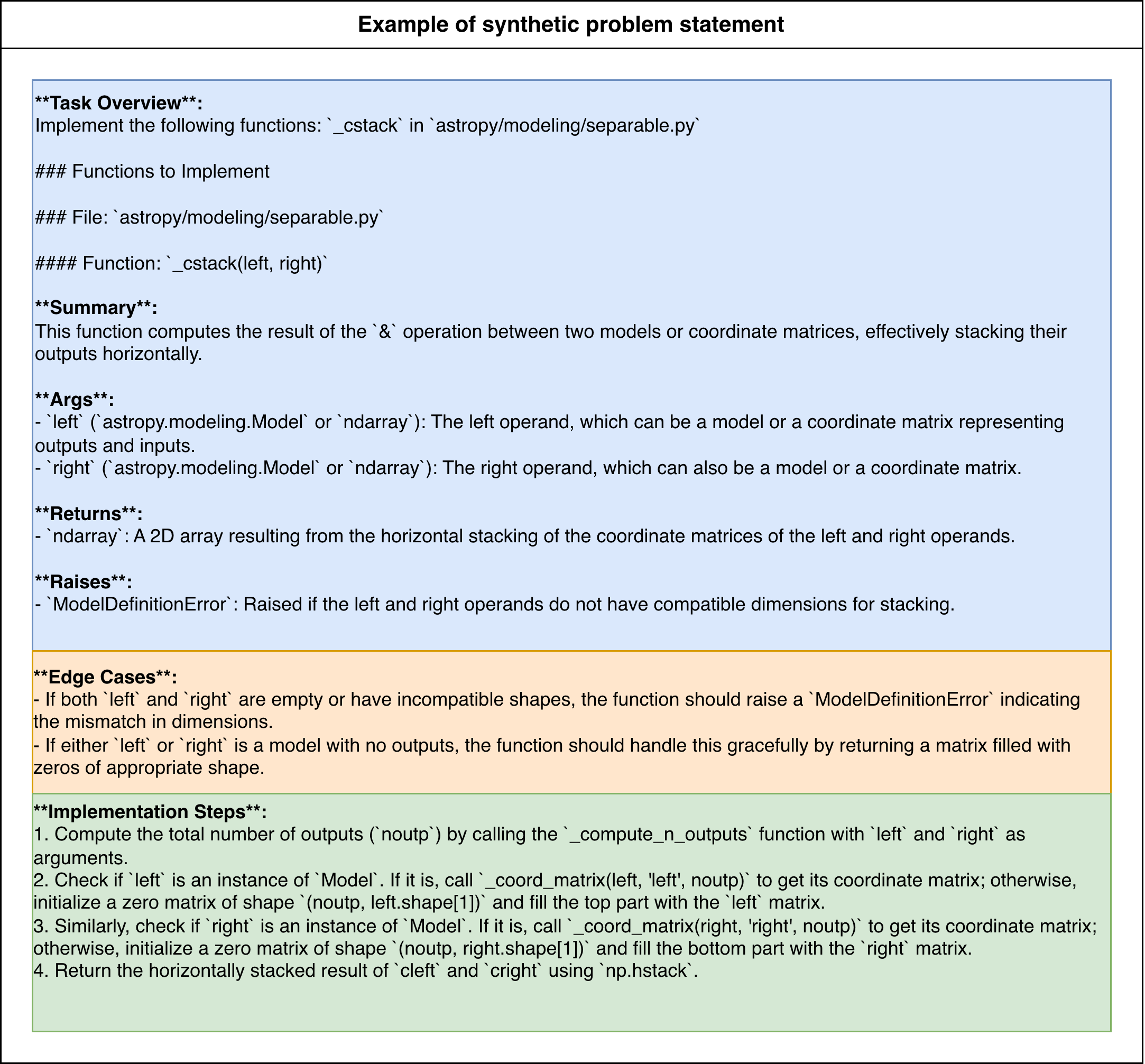} 
    \caption{\textbf{Prompt template used for solving $\text{PR}_1$.} The prompt is formatted in markdown and includes multiple components: a task overview, function summary, arguments, returns, raised exceptions, edge cases, and step-by-step implementation guidance.}
    \label{fig:pr_1_prompt_templates}
\end{figure*}

Figure \ref{fig:pr_1_ps_prompt} shows the prompt used to generate synthetic $\text{PR}_1$ problem statements. Figure \ref{fig:pr_1_prompt_templates} shows the prompt template used for solving $\text{PR}_1$.

\section{Resolve Rate and Maintainability Metric Comparison on the Full Dataset}\label{appendix:resolve_rate_full_data}
Table \ref{tab:pr2_resolve_chain} shows the comparison of PR$_2$ resolution rates under HA and AA conditions.

\begin{table*}[h]
\definecolor{negcolor}{rgb}{0.70,0.13,0.13}
\definecolor{poscolor}{rgb}{0.0,0.6,0.3}

\centering
\small
\begin{tabular}{l l c c c c}
\toprule
\textbf{Benchmark}
& \textbf{Model}
& \textbf{\#Instances}
& $\mathbf{PR_2}^{HA}$ \textbf{Resolve}
& $\mathbf{PR_2}^{AA}$ \textbf{Resolve}
& $\Delta$\\
\midrule

SWE-Bench Verified &
Claude
& 215
& 152 (70.7\%)
& 149 (69.3\%)
& \textcolor{negcolor}{-1.4} \\

& GPT
& 186
& 114 (61.3\%)
& 108 (58.1\%)
& \textcolor{negcolor}{-3.2} \\


& GLM
& 207
& 137 (66.2\%)
& 131 (63.3\%)
& \textcolor{negcolor}{-2.9} \\

& MiniMax
& 233
& 173 (74.25\%)
& 162 (69.53\%)
& \textcolor{negcolor}{-4.72} \\

\midrule
SWE-Bench Multilingual
& Claude
& 38
& 21 (55.3\%)
& 21 (55.3\%)
& 0.0 \\

& GPT
& 36
& 17 (47.2\%)
& 18 (50.0\%)
& \textcolor{poscolor}{2.8} \\

& GLM
& 58
& 40 (69.0\%)
& 32 (55.2\%)
& \textcolor{negcolor}{-13.8} \\

& MiniMax
& 86
& 54 (65.9\%)
& 52 (63.4\%)
& \textcolor{negcolor}{-2.5} \\

\midrule

SWE-Bench Pro
& Claude
& 313
& 141 (45.1\%)
& 116 (37.1\%)
& \textcolor{negcolor}{-8.0}\\

& GPT
& 280
& 109 (38.9\%)
& 111 (39.6\%)
& \textcolor{poscolor}{1.0} \\

& GLM
& 397
& 179 (45.1\%)
& 135 (34.0\%)
& \textcolor{negcolor}{-11.1} \\

& MiniMax
& 395
& 171 (42.3\%)
& 150 (38.0\%)
& \textcolor{negcolor}{-4.3} \\

\midrule

FeatBench
& Claude
& 17
& 8 (47.1\%)
& 9 (52.9\%)
& \textcolor{poscolor}{5.8} \\

& GPT
& 17
& 9 (52.9\%)
& 7 (41.2\%)
& \textcolor{negcolor}{-11.7} \\

& GLM
& 42
& 18 (42.9\%)
& 14 (33.3\%)
& \textcolor{negcolor}{-9.5} \\

& MiniMax
& 40
& 19 (47.5\%)
& 16 (40.0\%)
& \textcolor{negcolor}{-7.5} \\

\bottomrule
\end{tabular}
\caption{\textbf{PR$_2$ resolution rates under HA and AA conditions.} Across most model-benchmark pairs, building on agent code drops the resolve rate relative to building on human code. To save cost, the Claude and GPT-5 runs on SWE-bench Pro, Multilingual, and FeatBench are restricted to instances where the PR$_1$ gate passed for both MiniMax and GLM-4.7.}
\label{tab:pr2_resolve_chain}
\end{table*}




\section{Agent Execution and Model Details}\label{appendix:executor-details}

\subsection{Agent Execution}
All experiments use SWE-Agent under a single shared configuration: a \texttt{step\_limit} of 250 steps, a \texttt{cost\_limit} of \$3 USD per instance, and high reasoning effort across all models. For each agent-authored patch, we run the harness once to produce a patch. We adopt each benchmark's official evaluation harness, and for FeatBench we follow the harbor framework's open pull request (\href{https://github.com/harbor-framework/harbor/pull/1218}{\#1218}).

\subsection{Model link list}\label{appendix:model-link-list}

Table \ref{tab:model_links} presents the closed- and open-source models included in our benchmark.

\begin{table*}[h]
    \centering
    \small
    \begin{tabular}{l l l}
        \toprule
        \textbf{Model} & \textbf{Provider} & \textbf{Reference} \\
        \midrule
        \multicolumn{3}{l}{\textit{Closed-source models}} \\
        \midrule
        Claude 4.5 Sonnet & Anthropic & \url{https://www.anthropic.com/claude} \\
        GPT-5             & OpenAI    & \url{https://openai.com/index/introducing-gpt-5/} \\
        \midrule
        \multicolumn{3}{l}{\textit{Open-source models}} \\
        \midrule
        GLM 4.7 FP8       & Z.ai      & \url{https://huggingface.co/zai-org/GLM-4.7-FP8} \\
        MiniMax M2.5      & MiniMax   & \url{https://huggingface.co/MiniMaxAI/MiniMax-M2.5} \\
        \bottomrule
    \end{tabular}
    \caption{Models evaluated in this work, grouped by access type. Closed-source models are accessed via provider APIs; open-source models are run via self-hosted vLLM.}
    \label{tab:model_links}
\end{table*}

\section{LR Feature Details and LLM-Judge Details}\label{appendix:lr-feature-details}

\subsection{Maintainability Metric Computation}
We compute four maintainability metrics---cyclomatic complexity (CC), cognitive 
complexity (CogC), Halstead volume (HV), and logical lines of code (LLOC)---across 
all supported languages in our corpus. No single tool covers every metric across 
every language, so we combine multiple open-source libraries and one custom tool, 
selected to maximize per-language consistency. Table~\ref{tab:metric-libraries} 
lists each library's metric coverage. To fill a gap in Go tooling, we extend Maintidx, an existing Go library, into a custom \texttt{go-halstead} package that computes Halstead volume.

\begin{table*}[h]
    \centering
    \small
    \begin{tabular}{l l c c c c}
        \toprule
        \textbf{Library} & \textbf{Languages} & \textbf{CC} & \textbf{CogC} & \textbf{HV} & \textbf{LLOC} \\
        \midrule
        
        \href{https://pypi.org/project/radon/}{radon} & Python & \checkmark & — & \checkmark & \checkmark \\
        \href{https://pypi.org/project/complexipy/0.3.1/}{complexipy} & Python & — & \checkmark & — & — \\
        
        \href{https://github.com/mozilla/rust-code-analysis}{rust-code-analysis-cli} & Rust, C, Java, JS, TS & \checkmark & \checkmark & \checkmark & \checkmark \\
        
        go-halstead (custom) & Go & — & — & \checkmark & \checkmark \\
        \href{https://github.com/uudashr/gocognit}{gocognit} & Go & — & \checkmark & — & — \\
        \href{https://pypi.org/project/lizard/}{lizard} & Go, Ruby & \checkmark & — & — & \checkmark \\
        \bottomrule
    \end{tabular}
    \caption{Metric coverage of the libraries used to compute maintainability 
    metrics. \checkmark indicates supported, — indicates unsupported.}
    \label{tab:metric-libraries}
\end{table*}

\subsection{Patch Localization Features}\label{appendix:patch-localization}

The patch localization features measure where each chain's $\text{PR}_2$ patch edits. For each instance and each condition (HA, AA), we extract the set of files and the set of (file, function) pairs touched by the $\text{PR}_2$ patch. We parse each post-patch source file with an AST parser (Python's built-in \texttt{ast} module\footnote{\url{https://docs.python.org/3/library/ast.html}} for Python and \texttt{lizard}\footnote{\url{https://github.com/terryyin/lizard}} for all other supported languages) and compare against the pre-patch state to identify which files and functions the agent added, edited, or removed. We then compute the Jaccard overlap between the HA and AA edit sets on the same instance. Given two sets $A$ and $B$, the Jaccard overlap is:
  
\begin{equation*}
    J(A, B) = \frac{|A \cap B|}{|A \cup B|}.
\end{equation*}

For each discordant instance we compute two Jaccards: $J^{\text{file}}_i$ over the file sets and $J^{\text{func}}_i$ over the (file, function) sets, comparing HA's $\text{PR}_2$ patch with AA's $\text{PR}_2$ patch on the same instance. A Jaccard of 1.0 means the two chains edited exactly the same set; a Jaccard of 0 means the two sets are disjoint.

\subsection{$\text{PR}_1$ Behavioral Drift}\label{appendix:pr1-fidelity-judge}

\begin{figure*}[h]
     \centering
      \includegraphics[width=1.0\textwidth]{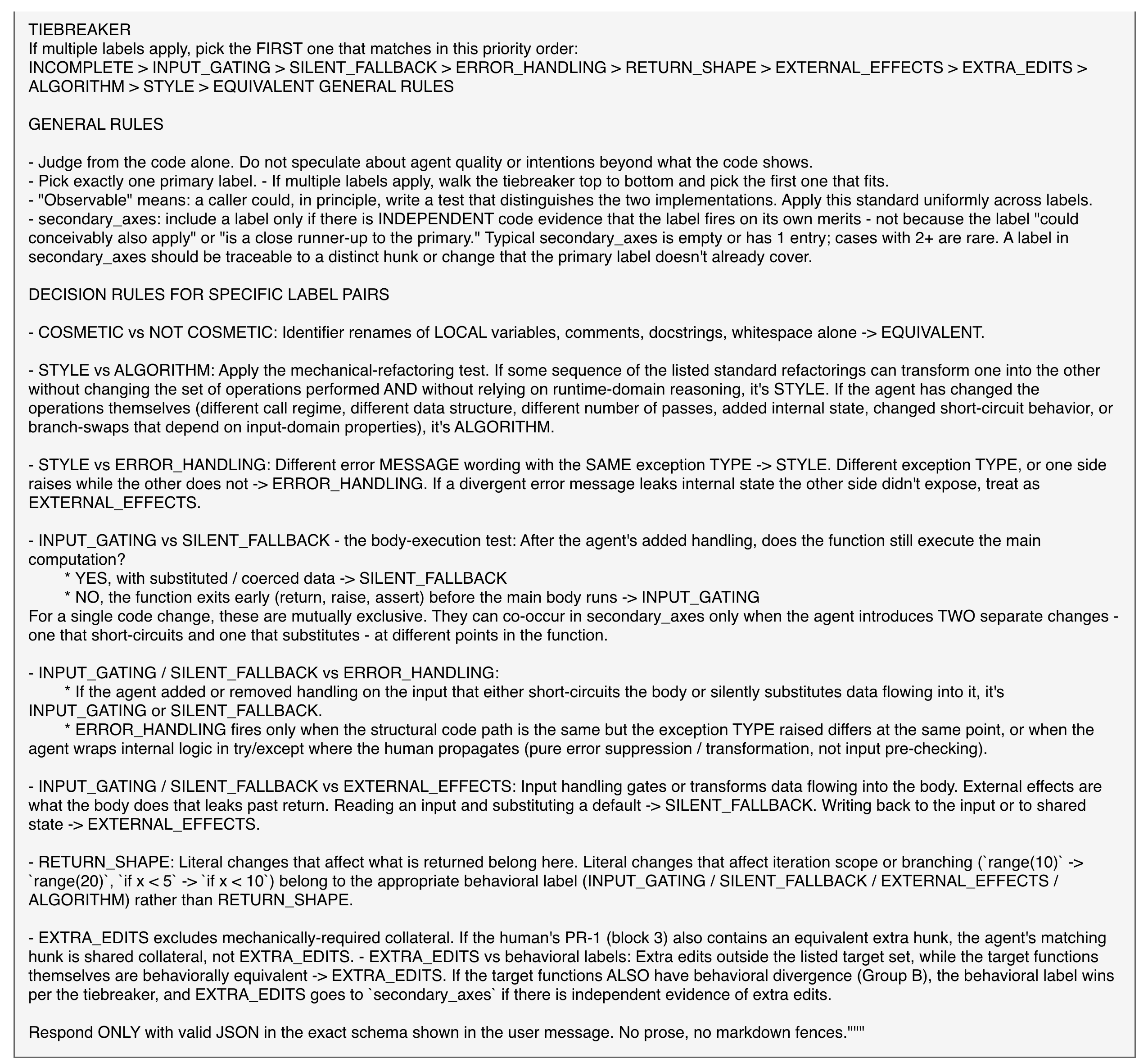}
    \caption{\textbf{System prompt for the $\text{PR}_1$ behavioral drift judge.} Defines the ten-category taxonomy.}
    \label{fig:llm-as-a-judge-pr1-fidelity-part-2}
\end{figure*}

\begin{figure*}[h]
     \centering
      \includegraphics[width=1.0\textwidth]{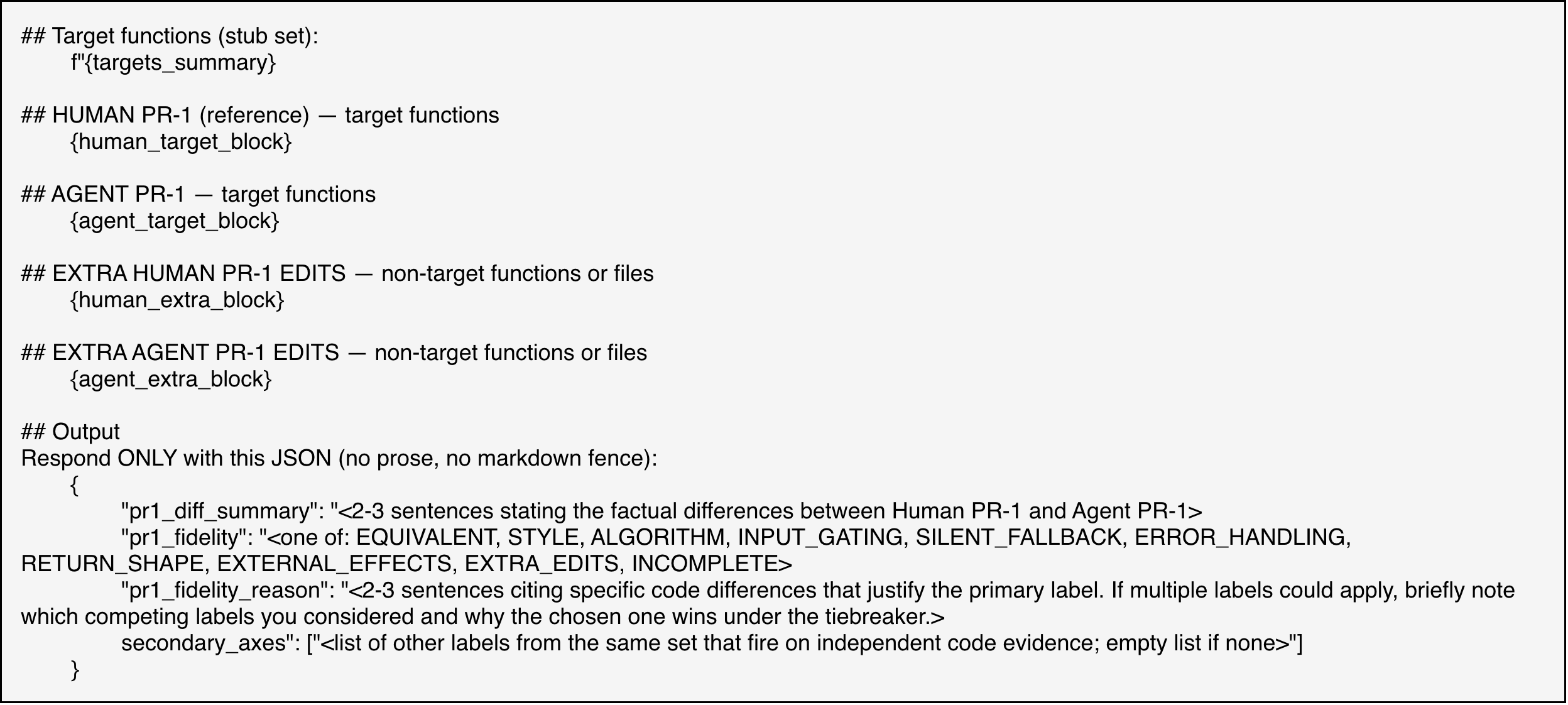} 
    \caption{\textbf{User prompt for the $\text{PR}_1$ behavioral drift judge.} Defines the output schema.}
    \label{fig:llm-as-a-judge-pr1-fidelity-user-prompt}
\end{figure*}

Prior work on agent failure analysis typically looks at the trajectory and the final agent-produced patch to classify failures into various categories~\cite{xu2025swecompassunifiedevaluationagentic, deng2025swebenchproaiagents}. Given our two-stage setup, we adapt this taxonomy to capture how Agent $\text{PR}_1$ diverges from Human $\text{PR}_1$. These divergences can silently propagate and cause failure during $\text{PR}_2$, as observed in Section~\ref{sec:analysis}. We develop a ten-category taxonomy that captures changes to the function's input contract, changes to its observable outputs and side effects, and edits that go beyond the stubbed target functions. Any of these can leave the downstream $\text{PR}_2$ task unresolved. Figure~\ref{fig:llm-as-a-judge-pr1-fidelity-part-2} shows the full system prompt. The prompt contains the definition of each category. We also reproduce the full definitions in Table~\ref{tab:pr1-fidelity-categories}. Figure~\ref{fig:llm-as-a-judge-pr1-fidelity-user-prompt} shows the user prompt and the output schema. We use DeepSeek V4 Pro as our LLM judge. We run it on 2{,}517 records, one for every instance in our four-model by four-benchmark grid.

\subsection{$\text{PR}_1$ Error Compounding into $\text{PR}_2$} \label{appendix:pr2-pr1-judge}

\begin{figure*}[h]
     \centering
      \includegraphics[width=0.95\textwidth]{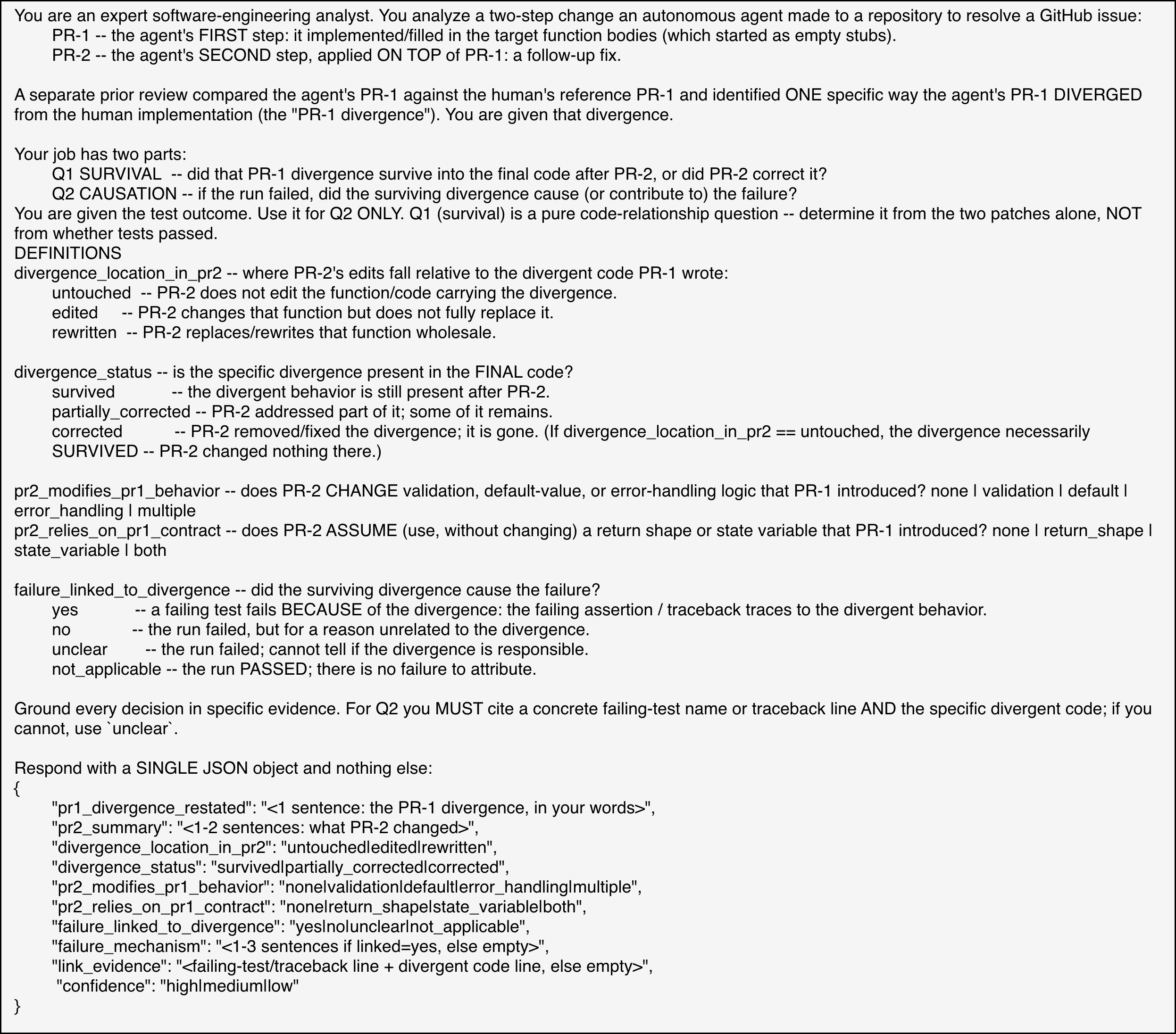}
    \caption{\textbf{System prompt for the $\text{PR}_1$/$\text{PR}_2$ compounding judge.}}
    \label{fig:pr2-pr1-system-prompt}
\end{figure*}

The $\text{PR}_1$ behavioral drift judge in Appendix~\ref{appendix:pr1-fidelity-judge} identifies instances where Agent $\text{PR}_1$ diverges from Human $\text{PR}_1$. A $\text{PR}_1$ divergence does not automatically mean the chain fails. Two scenarios can decouple it from the dwonstream failure. First, the agent's $\text{PR}_2$ may rewrite the divergent function, replacing the drift with a different code block. Second, even when the drift survives, the benchmark's unit tests may not exercise it. The question is therefore: did a $\text{PR}_1$ divergence persist into $\text{PR}_2$'s final state, and did it directly relate to the failing test case? We use DeepSeek-V4 Pro to judge AA instances. Figure~\ref{fig:pr2-pr1-system-prompt} and Figure~\ref{fig:pr2-pr1-user-prompt} show the full system and user prompt with the output schema. We exclude instances labeled \texttt{EQUIVALENT} or \texttt{STYLE} by the $\text{PR}_1$ behavioral drift judge because they have no divergence to compound, which leaves 2{,}153 instances.

\begin{table*}[h]
      \centering
      \small
      \setlength{\tabcolsep}{6pt}
      \begin{minipage}[t]{0.48\linewidth}
          \centering
          \begin{tabular}{l r r}
              \toprule
              \textbf{Outcome} & \textbf{n} & \textbf{\%} \\
              \midrule
              Drift survived         & 225 & 85.9 \\
              Partially corrected    &  19 &  7.3 \\
              Rewrote / corrected    &  18 &  6.9 \\
              \bottomrule
          \end{tabular}
          \caption*{\textbf{Did the $\text{PR}_1$ drift survive $\text{PR}_2$?} $\text{PR}_2$ leaves the agent's $\text{PR}_1$ drift unchanged on 85.9\% of the 262 discordant instances, suggesting that the drift almost always persists into $\text{PR}_2$.}
      \end{minipage}
      \hfill
      \begin{minipage}[t]{0.48\linewidth}
          \centering
          
          \begin{tabular}{l r r}
              \toprule
              \textbf{Outcome} & \textbf{n} & \textbf{\%} \\
              \midrule
              Yes (compounding-confirmed) & \textbf{54} & \textbf{20.6} \\
              No (unrelated failure)      & 195 & 74.4 \\
              Unclear                     &  12 &  4.6 \\
              Not applicable (passed)     &   1 &  0.4 \\
              \bottomrule
          \end{tabular}
          \caption*{\textbf{Did the failure trace back to the drift?} The surviving PR$_1$ drift directly causes the failing test on 20.6\% of the 262 discordant instances. The remaining instances fail for undetermined reasons.}
      \end{minipage}
      \caption{\textbf{LLM-judge attribution on the HA-wins discordant cohort} (n=262 instances with a divergent agent PR$_1$; 30 PR$_1$-equivalent/style cases excluded). Drift survives PR$_2$ in $86\%$ of cases; on $20.6\%$ of the
  cohort the surviving drift directly traces to the failing test.}
      \label{tab:pr1-drift-attribution}
  \end{table*}

\begin{table*}[h]
  \centering

  \begin{tabular}{lp{0.7\linewidth}}
  \toprule
  \textbf{Category} & \textbf{Definition} \\
  \midrule
  \multicolumn{2}{l}{Group A: functionally compatible} \\
  \midrule
  {EQUIVALENT} &
  Same logic, control flow, and observable contract. Only cosmetic differences such as whitespace, comments, docstrings, or local-variable renames. \\
  {STYLE} &
  Same observable contract and the same operations in the same order, expressed in a different surface syntactic form (e.g., list comprehension vs.\ loop, ternary vs.\ if/else, early return vs.\ nested). \\
  {ALGORITHM} &
  Same observable contract, but the operations themselves differ in identity, count, or resource regime (e.g., iterative vs.\ recursive, added memoization, different short-circuit ordering). \\
  \midrule
  \multicolumn{2}{l}{Group B: behavioral divergence} \\
  \midrule
  {INPUT GATING} &
  Agent added or removed an explicit input check that prevents the main body from executing on some inputs. The function exits early via a raise, an early return, or an assert. \\
  {SILENT FALLBACK} &
  Agent silently substituted or coerced an input so that the main body still runs but on different data (e.g., \texttt{dict.get(k, default)}, \texttt{int(x)}, \texttt{x or default}). \\
  {ERROR HANDLING} &
  Agent raises, catches, or wraps exceptions differently on a structurally equivalent code path (e.g., different exception type at the same raise site, propagating vs.\ catching). \\
  {RETURN SHAPE} &
  Return value's type, structure, or layout differs (e.g., \texttt{None} vs.\ \texttt{[]}, scalar vs.\ tuple, different dict keys or layout). \\
  {EXTERNAL EFFECTS} &
  Observable mutations or I/O differ (e.g., stdout or log calls added or removed, mutation of arguments, writes to instance or module state). \\
  \midrule
  \multicolumn{2}{l}{{Group C: scope}} \\
  \midrule
  {EXTRA EDITS} &
  Target functions are behaviorally compatible, but the agent also modified non-target files or functions that were not mechanically required. Imports and helpers called from the target body do not count. \\
  \midrule
  \multicolumn{2}{l}{{Group D: agent failure}} \\
  \midrule
  {INCOMPLETE} &
  Agent left at least one target stub in a non-functional or placeholder state (e.g., empty body, \texttt{raise NotImplementedError}, return paired with a \texttt{TODO} comment). \\
  \bottomrule
  \end{tabular}
  \caption{The ten $\text{PR}_1$ behavioral drift categories. The judge picks exactly one primary category per instance.}\label{tab:pr1-fidelity-categories}
\end{table*}

\begin{figure*}[h]
     \centering
      \includegraphics[width=1.0\textwidth]{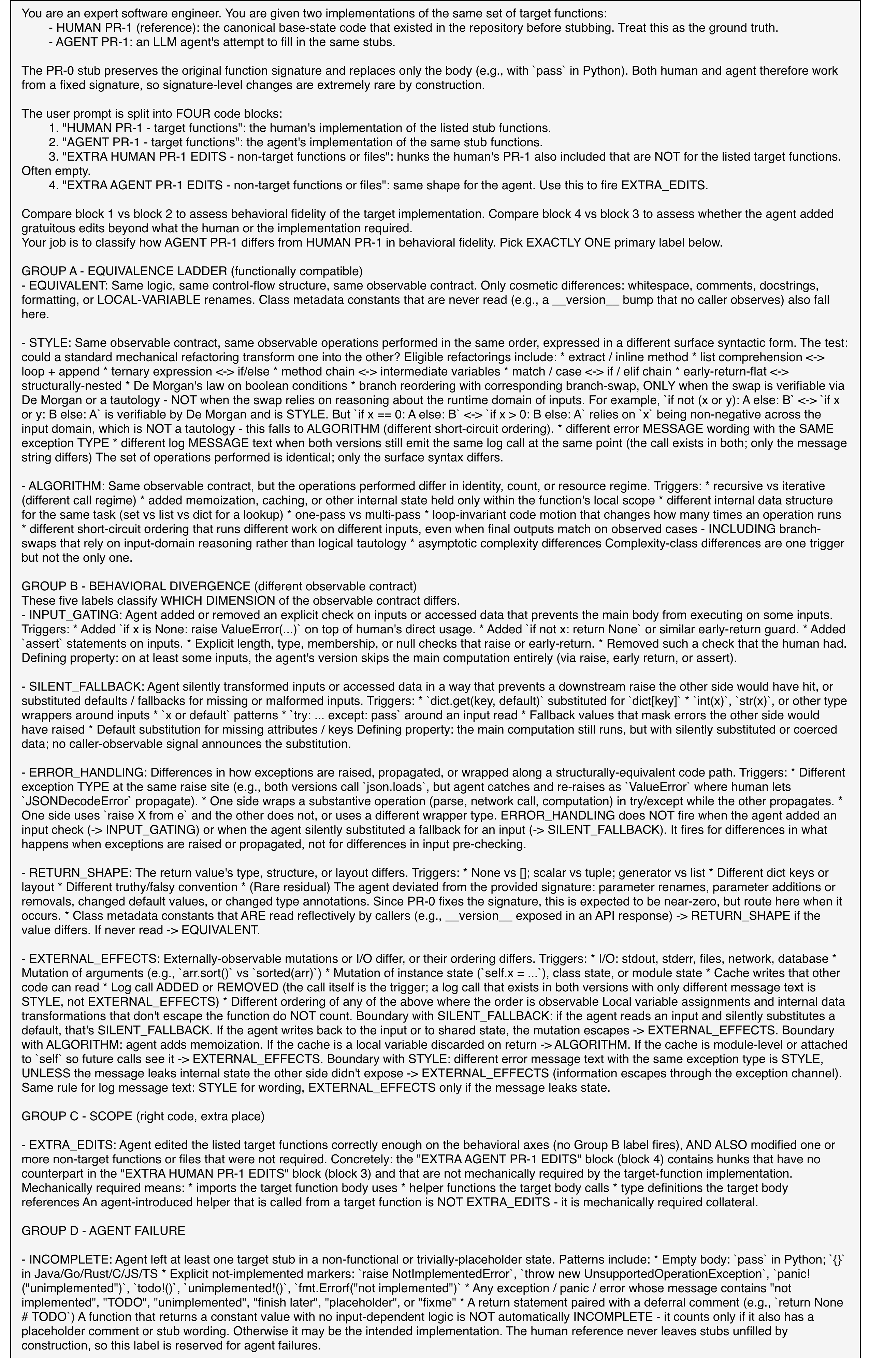} 
    \label{fig:llm-as-a-judge-pr1-fidelity-part-1}
\end{figure*}

\begin{figure*}[h]
     \centering
      \includegraphics[width=0.9\textwidth]{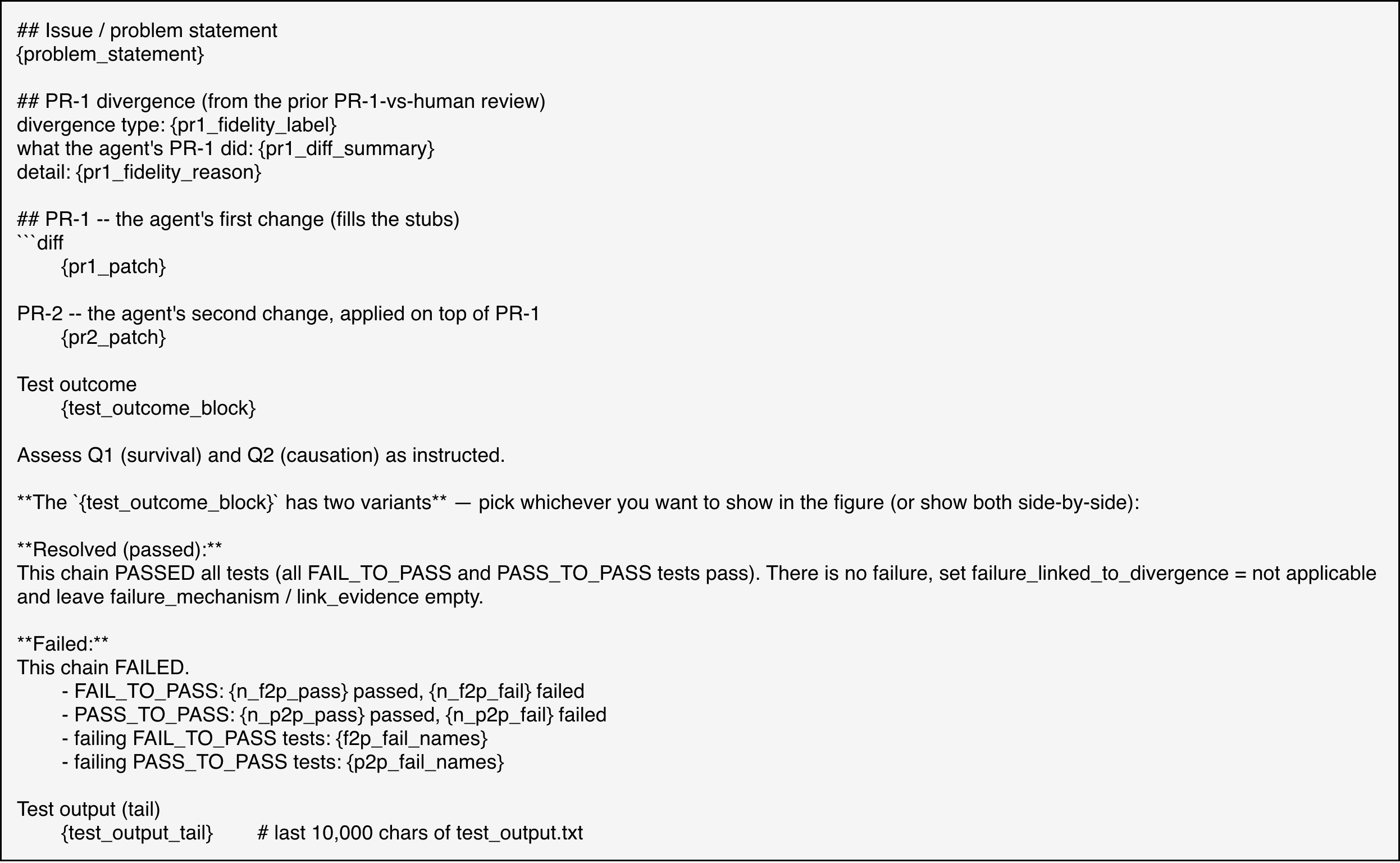} 
    \caption{\textbf{User prompt for the $\text{PR}_1$/$\text{PR}_2$ compounding judge} Defines the output schema.}\label{fig:pr2-pr1-user-prompt}
\end{figure*}

\subsection{Human Audit of the $\text{PR}_1$ Fidelity Judge}\label{appendix:human-audit}

We verify a stratified sample of the DeepSeek-V4 Pro judge's outputs against the underlying code to assess the reliability of its IEC labels. We sampled 50 instances, with 56\% labeled IEC and the remainder assigned non-IEC labels. Instances used as examples or illustrations elsewhere in the paper were excluded from the sample. Rare labels were over-sampled so that every label is represented. The sample spans all four models and all four benchmarks used in our evaluation. Samples range from 28 to 376 combined patch lines (median 109).

For each instance, the auditor was shown the judge's label and reasoning together with the human and agent $\text{PR}_1$ implementations of the same stubbed target functions. The auditor then determined whether the assigned label was correct using the same criteria specified in the judge prompt. We audit the judge's labels rather than obtain independent blind annotations given the difficulty of the task---assigning a label from scratch often requires repository-specific knowledge beyond the patch itself, such as whether an attribute may take a false value, and the sample of 50 instances covers 29 distinct repositories.
As such, this procedure is a verification audit rather than an independent re-annotation, and we report confirmation rates.

The IEC/non-IEC classification underlying our analysis was confirmed in 48 of 50 instances (96.0\%). Seven of these moved an item between two labels on the same side of the boundary, leaving its IEC classification unchanged. The remaining two were reclassified from IEC to non-IEC, and no item was reclassified from non-IEC to IEC.

\subsection{Robustness to Choice of LLM Judge}\label{appendix:judge-robustness}
  
IEC is the only feature in our model that comes from an LLM judge. To check that the result does not depend on a judge model, we re-ran the $\text{PR}_1$ behavioral drift judge with Kimi K2.6 on the $405$ discordant instances used to fit the model in Section~\ref{sec:analysis}. We rebuilt the IEC indicator from its labels and re-fit the same LR model by keeping everything constant. Table~\ref{tab:judge-robustness} compares the significant coefficients against Table~\ref{tab:disagreement-lr}; all three keep their sign, their significance, and their approximate magnitude.
      
\begin{table}[t]
  \centering
  \small
  \setlength{\tabcolsep}{6pt}
  \begin{tabular}{lcccc}
      \toprule
      & \multicolumn{2}{c}{\textbf{DeepSeek V4 Pro}} & \multicolumn{2}{c}{\textbf{Kimi K2.6}} \\
      \cmidrule(lr){2-3}\cmidrule(lr){4-5}
      \textbf{Feature} & \textbf{OR} & $\boldsymbol{p}$ & \textbf{OR} & $\boldsymbol{p}$ \\
      \midrule                          
      $\Delta\text{LLOC}_{\text{PR}_2}$ & $1.88$ & $0.042$ & $1.83$ & $0.042$ \\
      IEC                               & $1.83$ & $0.011$ & $1.70$ & $0.039$ \\
      Inst.\ resolve rate               & $1.42$ & $0.003$ & $1.42$ & $0.003$ \\
      \bottomrule
  \end{tabular}
  \caption{\textbf{Swapping the LLM judge does not change our findings.} We re-fit the model in Table~\ref{tab:disagreement-lr} using IEC labels from a second judge, Kimi K2.6. The same three features stay significant, with similar odds ratios.}
  \label{tab:judge-robustness}
\end{table}

\subsection{Examples of IEC Divergence} \label{sec:iec_examples} 
\begin{figure*}[h]
     \centering
       \includegraphics[width=\textwidth]{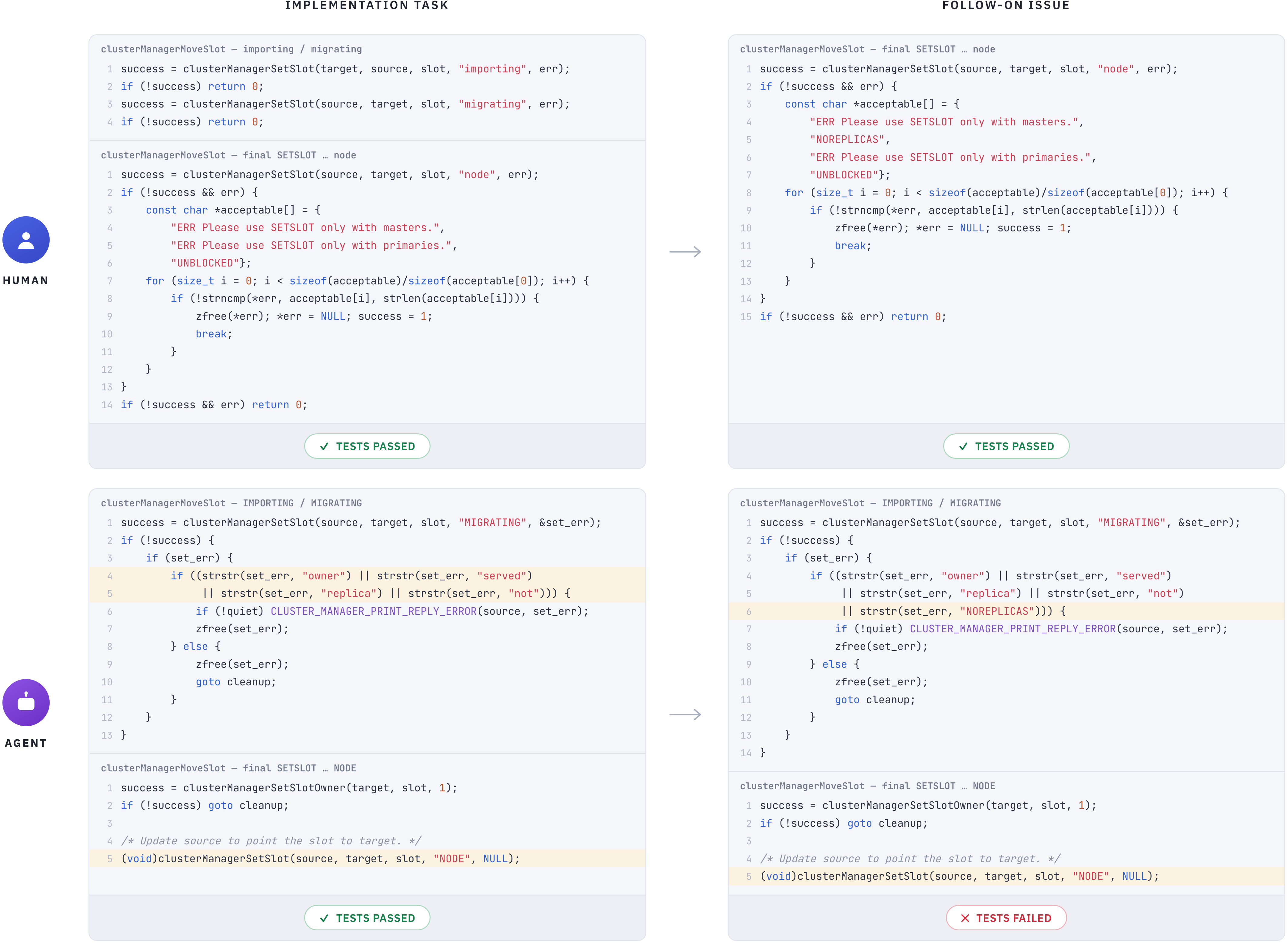} 
    \caption{\textbf{Example of a silent fallback.} In human $\text{PR}_1$, \texttt{clusterManagerMoveSlot} returns zero on failure (line 4), propagating the failure to its caller. In agent $\text{PR}_1$ (lines 2--13), a failed call is handled differently: the code checks whether the error message contains certain keywords and, if so, prints the error and continues execution instead of returning zero. This behavior persists into agent $\text{PR}_2$ and causes the follow-up test to failure.}
\label{fig:pr2-pr1-user-prompt}\label{fig:iec_s1}
\end{figure*}

\begin{figure*}[h]
     \centering
      \includegraphics[width=\textwidth]{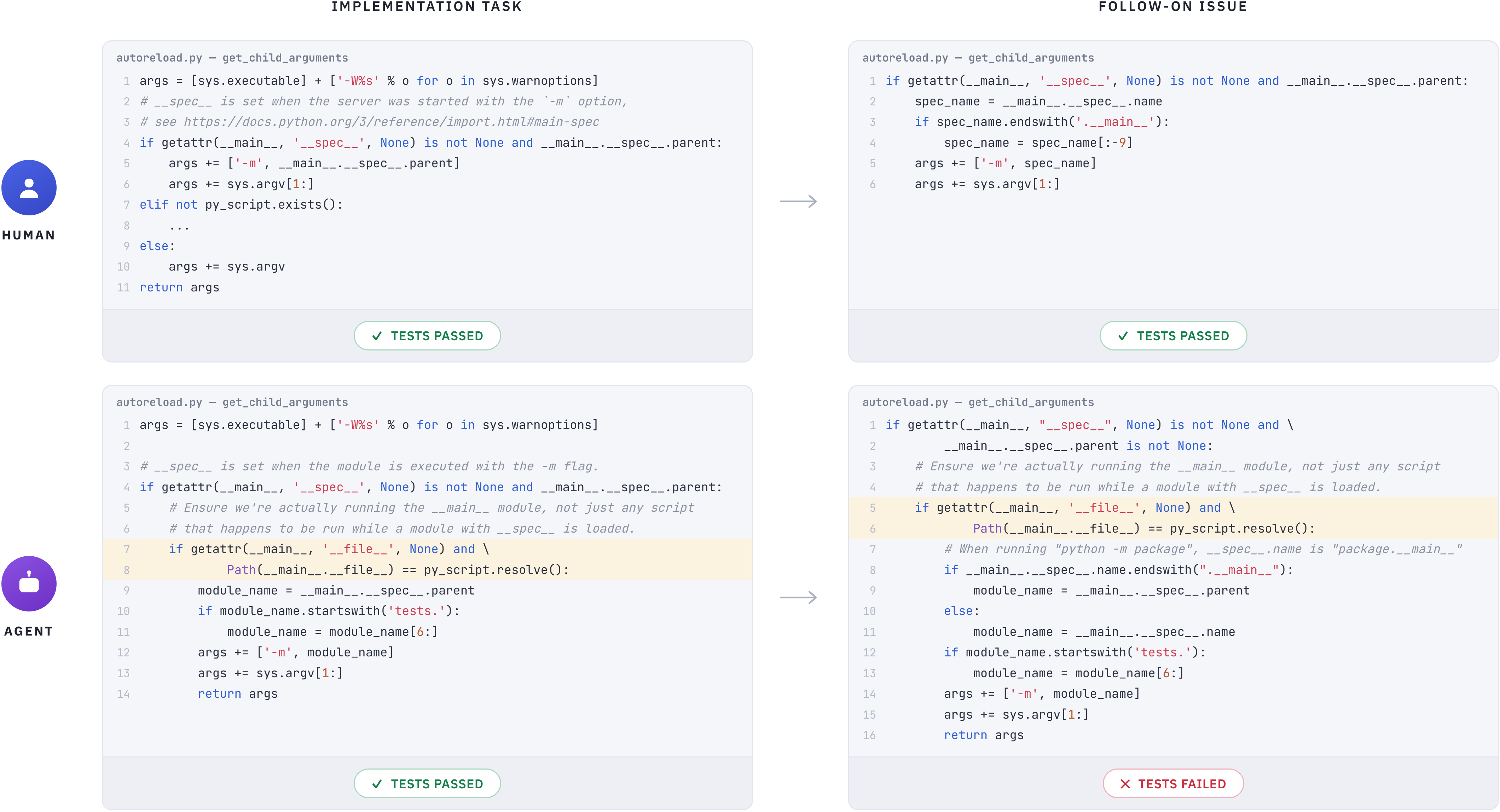} 
    \caption{\textbf{Example of added input gating.} In Agent $\text{PR}_1$, the condition on line 5 introduces an additional input check. This condition is carried over into Agent $\text{PR}_2$, where it causes the follow-up test to fail.
    }\label{fig:pr2-pr1-user-prompt}\label{fig:iec_s2}
\end{figure*}

\begin{figure*}[h]
     \centering
      \includegraphics[width=\textwidth]{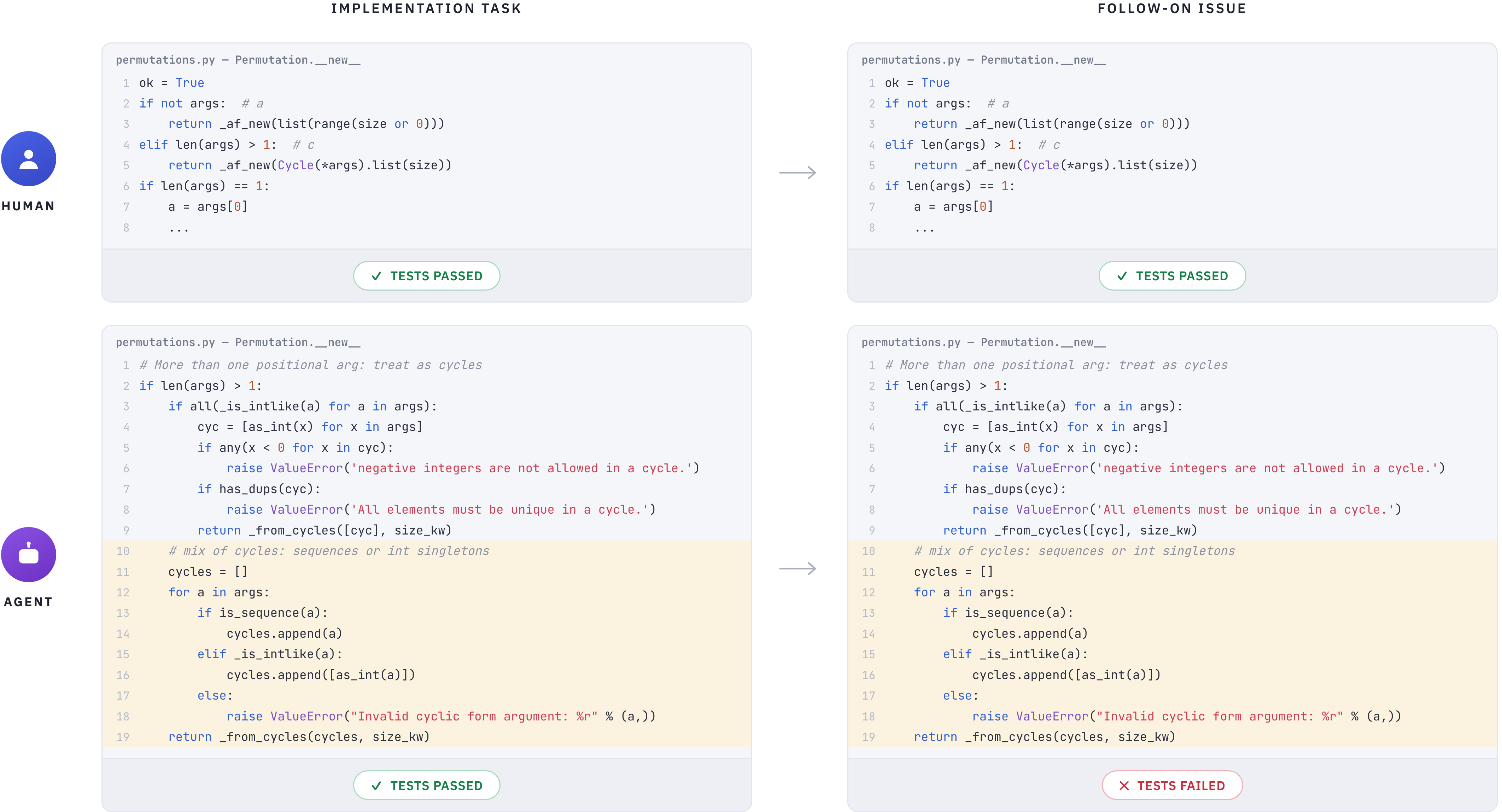} 
    \caption{\textbf{Example of removed input gating.} The human authored  $\text{PR}_1$ rejects a call made with two list arguments. The agent's $\text{PR}_1$ accepts them and builds a result instead (highlighted). $\text{PR}_2$  edits another function, leaving this behaviour exactly as $\text{PR}_2$ left it, so the test that expects the call to be rejected fails.}\label{fig:iec_s3}
\end{figure*}

\begin{figure*}[h]
     \centering
      \includegraphics[width=\textwidth]{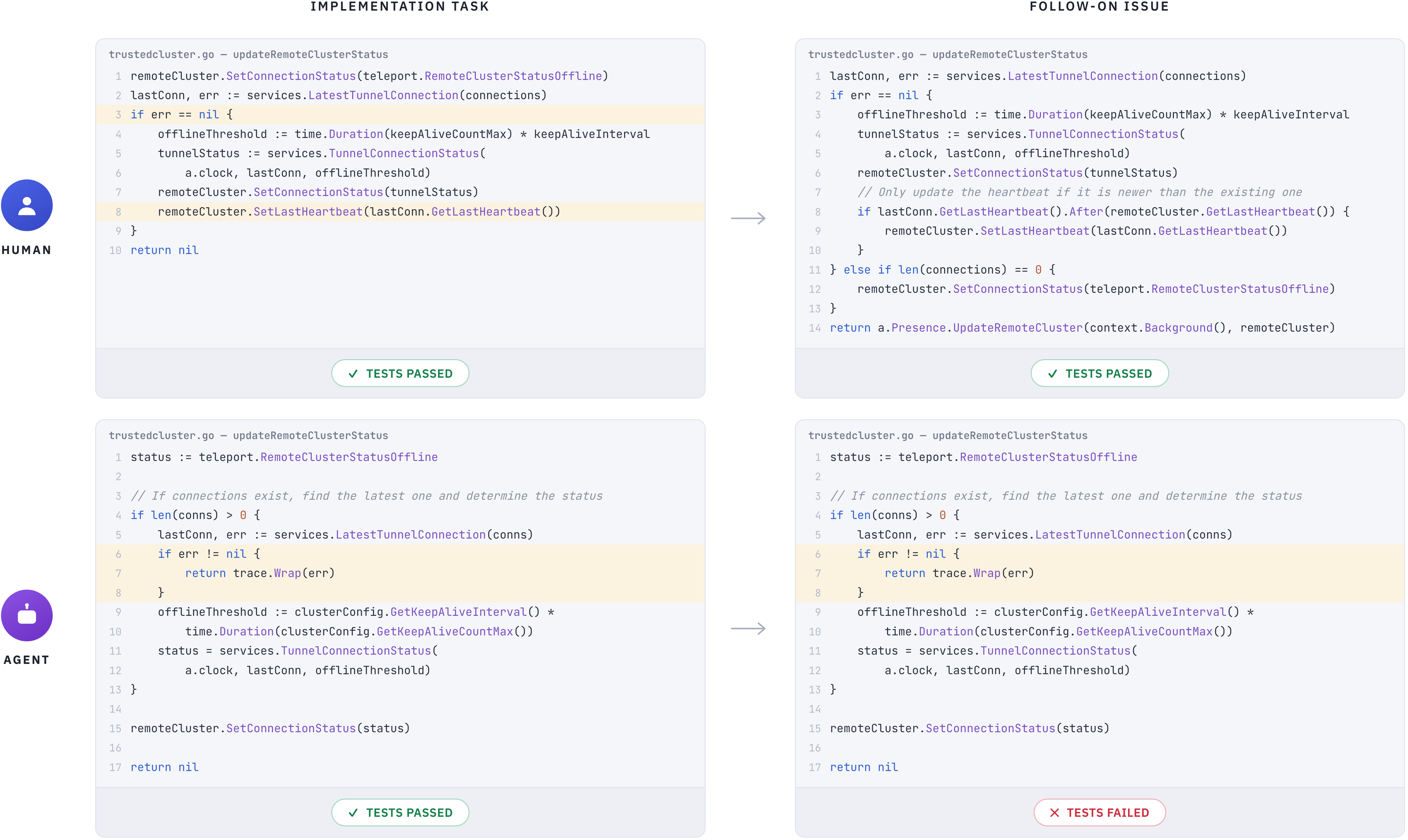} 
    \caption{\textbf{Example of changed error handling.} When \texttt{err} is \texttt{nil}, human $\text{PR}_1$ records and stores the latest heartbeat. In agent $\text{PR}_1$, the heartbeat is not stored; instead, the function returns to its caller. This behavior is preserved in Agent $\text{PR}_2$, which returns an empty heartbeat date and causes the follow-up test to fail.}\label{fig:iec_s4}
\end{figure*}

\begin{figure*}[h]
     \centering
      \includegraphics[width=\textwidth]{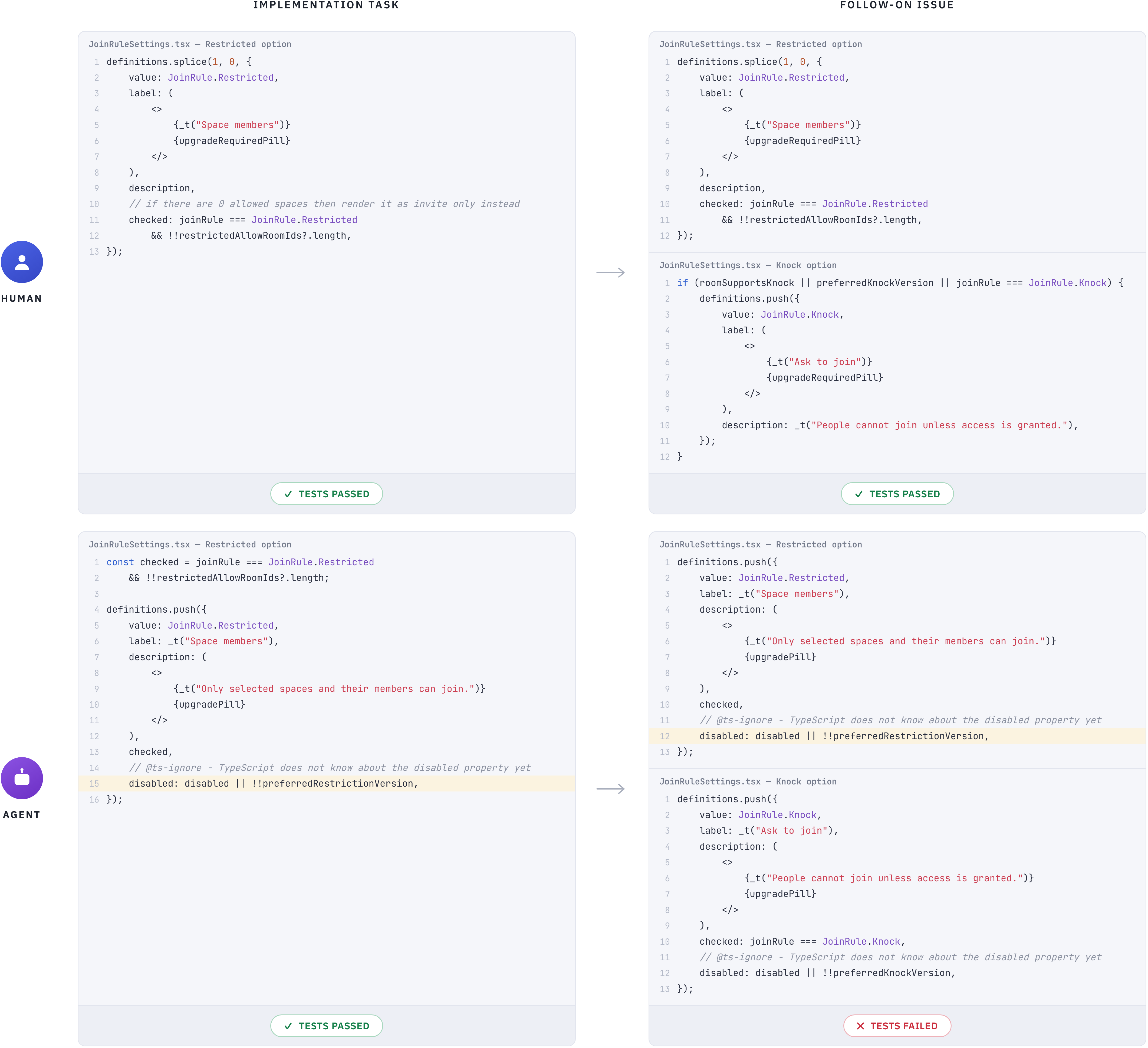} 
\caption{\textbf{Example of added input gating.} Human $\text{PR}_1$ does not include a \texttt{disabled} property, while Agent $\text{PR}_1$ adds one. This changes the frontend behavior: clicking the option no longer opens the dialog box. The same behavior persists in Agent $\text{PR}_2$ and is also applied to a newly added option. As a result, all nine follow-up tests fail because the expected dialog never appears.}\label{fig:iec_s5}
\end{figure*}

Each figure below shows an instance with the human and agent implementations of the same function after PR1 on the left, and what each becomes after the same follow-on issue is addressed on the right. Shaded lines mark the code identified as divergent. The five span all three divergence types and four languages: Figures~\ref{fig:iec_s1} and~\ref{fig:iec_s4} in C and Go, Figures~\ref{fig:iec_s2} and~\ref{fig:iec_s3} in Python, and Figure~\ref{fig:iec_s5} in TypeScript.

\end{document}